\documentclass[a4paper,11pt]{article}
\pdfoutput=0
\usepackage{jheppub}

\usepackage{epsfig,afterpage}
\usepackage{graphicx}
\usepackage{amsfonts}
\usepackage{amssymb}
\usepackage{indentfirst}
\usepackage{amsmath,amsthm}
\usepackage{dsfont}
\usepackage{slashed} 

\usepackage{epsfig}
\usepackage{subfigure}

\usepackage{multirow}
\usepackage{pstricks}
\usepackage{psfrag}
\usepackage{lscape}

\usepackage{wasysym} 

\usepackage{algorithm,algorithmic,caption}

\def\beq{\begin{equation}}
\def\eeq{\end{equation}}
\def\bea{\begin{eqnarray}}
\def\eea{\end{eqnarray}}
\def\bq{\begin{quote}}
\def\eq{\end{quote}}
\def\ben{\begin{enumerate}}
\def\een{\end{enumerate}}
\def\bit{\begin{itemize}}
\def\eit{\end{itemize}}

\notoc

\begin{document}

\title{The mass spectrum of the Schwinger model with Matrix Product States}

 \author[a]{M. C. Ba\~nuls,}
 \emailAdd{banulsm@mpq.mpg.de} 
 \affiliation[a]{Max-Planck-Institut f\"ur Quantenoptik,
 Hans-Kopfermann-Str. 1, 85748 Garching, Germany}
 \author[b,c]{K. Cichy,} 
 \affiliation[b]{NIC, DESY Zeuthen, Platanenallee 6, 15738 Zeuthen, Germany}
 \affiliation[c]{Adam Mickiewicz University, Faculty
of Physics,
Umultowska 85, 61-614 Poznan, Poland}
\author[a]{J. I. Cirac} 
\author[b,d]{and K. Jansen} 
\affiliation[d]{Department of Physics, University of Cyprus, P.O. Box 20537, 1678 Nicosia,
Cyprus}


\abstract{
We show the feasibility of tensor network solutions for lattice gauge theories in Hamiltonian formulation by applying 
matrix product states algorithms to the Schwinger model with zero and
non-vanishing fermion mass.
We introduce new techniques to compute excitations in a system with open boundary conditions, and to 
identify the states corresponding to low momentum and different quantum numbers in the continuum.
For the ground state and both the vector and scalar mass gaps in the massive case, the MPS technique 
attains precisions comparable to the best results available from other techniques.
}

\preprint{DESY 13-084, SFB/CPP-13-31}


\maketitle

\section{Introduction}
\label{sec:intro}

In the last years, methods based on tensor network states (TNS) have revealed themselves as very promising tools 
for the numerical study of strongly correlated quantum many-body systems.
They are ans\"atze for the quantum
state,
characterized by their entanglement content and well suited for lattice systems. 
The paradigmatic family of TNS is that of matrix product states (MPS) \cite{aklt88,kluemper91,kluemper92,fannes92fcs,verstraete04dmrg,perez07mps}, 
which underlies the well-known 
density matrix renormalization group algorithm (DMRG) \cite{white92dmrg,schollwoeck05dmrg}.
The enormous success of DMRG for the study of one dimensional condensed matter problems
has been better understood  within the framework of TNS (see e.g. \cite{schollwoeck11age}).
This has also enabled different extensions of the original algorithm, such as time-dependence \cite{vidal03eff,daley04adapt,vidal07infinite}, so that
MPS/DMRG methods constitute nowadays a quasi exact method for the study of ground states, low lying excitations and thermal equilibrium 
properties of quantum spin chains far beyond the reach of exact diagonalization.
Being free from the sign problem which plagues quantum Monte Carlo (QMC) methods, higher dimensional
TNS \cite{verstraete04peps,vidal07mera,cirac09rg} are
seen as powerful candidates for the numerical exploration of long standing strongly correlated electron problems.


This success has motivated the application of TNS techniques to further problems, 
and a relatively new field of application has been found in quantum field theories.
Conformal Field Theory inspires a generalization of MPS \cite{cirac10iMPS}
useful for critical models.
Furthermore, specific generalizations of TNS
exist \cite{verstraete10cMPS,osborne10holo,haegeman13cMERA}
which are suitable for non-relativistic and relativistic QFT in the continuum.
The discrete versions, on the other hand, are adequate for 
lattice field theories, and can in particular be applied to 
models relevant to high energy physics problems.
This route was first explored using the original DMRG formulation~ \cite{byrnes02dmrg,sugihara05gaugeinv}.
The deeper understanding of the technique achieved thanks to TNS has increased the power of 
these methods \cite{tagliacozzo11} and new results are now available for critical QFT \cite{weir10,milsted13mpsqft}.

A particularly interesting study case is that of the Schwinger model \cite{schwinger62,coleman76},
or QED in one spatial dimension,
the simplest gauge theory, which nevertheless exhibits features 
in common with more complex models (QCD) such as confinement or a non-trivial vacuum,
and has been adopted as a benchmark model where to explore lattice gauge theory
techniques (see
e.g. \cite{Gutsfeld:1999pu,Gattringer:1999gt,Giusti:2001cn,Christian:2005yp,Bietenholz:2011ey}
and references cited therein).

The application of MPS techniques to the Schwinger model was first explored by Byrnes et
al. \cite{byrnes02dmrg}
using the original DMRG formulation. The study improved by several orders of magnitude the precision
attained by other Hamiltonian techniques for the ground state and the first (vector) mass gap,
for vanishing and non-vanishing fermion masses, using periodic boundary conditions (PBC).
However, DMRG estimations for higher excited states lose precision fast, in particular for PBC,
and the scalar mass gap was not explored in \cite{byrnes02dmrg}.

In this paper we apply the more stable and numerically efficient MPS for open boundary conditions 
(OBC) to the same problem. 
We work in the subspace of physical states satisfying Gauss' law, in which the model can be written
as a spin Hamiltonian with a long-range
interaction \cite{Banks:1975gq,crewther80}. 
We improve the existing techniques to find higher excited states 
and devise a method to identify vector and scalar excitations in finite open chains, although the charge 
that distinguishes them is only a good quantum number in the continuum.
Using these methods we compute  the ground state and the vector and scalar 
mass gaps for vanishing and non-vanishing fermion masses, 
with enough precision to conduct the extrapolation to the continuum limit.
Our study shows the feasibility of TNS solutions for similar LGT,
in the Hamiltonian formulation.

The rest of the paper is organized as follows.
In section~\ref{sec:model} we briefly introduce the Schwinger model, and review its formulation
as a spin Hamiltonian. Section~\ref{sec:method} presents the MPS formalism, 
with particular emphasis on the new techniques used to
overcome the challenge posed by this particular kind of problems. 
In section~\ref{sec:results} we present our numerical results and conclude in section~\ref{sec:discussion} with a discussion and outlook.

\section{The model}
\label{sec:model}

The massive Schwinger model \cite{schwinger62}, or QED in two space-time dimensions, is a
gauge theory, describing the interaction of a single fermionic species
with the photon field, via the Lagrangian density,
\beq
{\cal{L}}=\bar{\Psi} (i \slashed{\partial}- g \slashed{A }-m )\Psi-\frac{1}{4}F_{\mu\nu}F^{\mu \nu},
\label{eq:lagrangian}
\eeq
with $F_{\mu\nu}=\partial_{\mu}A_{\nu}-\partial_{\nu}A_{\mu}$, $g$ is the coupling constant and $m$
the fermion mass.
The physics of the model is determined by the only dimensionless parameter $m/g$.
The massless case, $m=0$, can be solved analytically, via bosonization \cite{schwinger62},
and also the free case, $g=0$, has exact solution, so that for very large or very small values of 
$m/g$ a perturbative study is possible around one of the solvable limits, 
while in general a non-perturbative treatment may be needed.
One of the features of the model is the existence of bound states,
 the two lowest ones of which, the vector and the scalar, are stable for any value of $m/g$ 
\cite{coleman76,adam97}. 

The spectrum of the Schwinger model has been studied with many techniques. The most accurate numerical estimations 
have been performed on the lattice.
For the vector mass the best results were obtained with DMRG \cite{byrnes02dmrg}, 
while strong coupling expansion (SCE) got the most accurate prediction for the scalar mass \cite{hamer82,sriganesh2000}
and the most precise values in the massless case \cite{cichy13}.

In the temporal gauge, $A_0=0$, the Hamiltonian density reads
\beq
{\cal{H}}=-i\bar{\Psi}\gamma^1(\partial_1-igA_1)\Psi+m\bar{\Psi}{\Psi}+\frac{1}{2}E^2.
\eeq
The electric field, $E=-\dot{A}^1$, is fixed by the additional constraint of Gauss' law,
$\partial_1E=g\bar{\Psi}\gamma^0\Psi$, up to a constant of integration, which 
can be interpreted as a background field~Ê\cite{coleman76}.

The model can be formulated on a lattice.
Here we focus on
 the Kogut-Susskind staggered formulation \cite{kogut75sc},
\begin{equation}
H=-\frac{i}{2 a}\sum_n \left (  \phi^{\dagger}_n e^{i\theta_n}  \phi_{n+1} - \mathrm{h.c.}\right ) 
+m\sum_n(-1)^n\phi_n^{\dagger}\phi_n
+\frac{a g^2}{2} \sum_n L_n^2,
\label{eq:Hferm}
\end{equation}
where $a$ is the lattice spacing.
We consider a lattice with open boundaries, with sites $n=0,\ldots N-1$.
On each site there is a single-component fermion field $\phi_n$, while $\theta_n$ are
the gauge variables sitting on the links between $n$ and $n+1$, 
and connected to the vector potential via $\theta_n=-a q A_n^1$.
$L_n$ is the corresponding conjugate variable, $[\theta_n,L_m]=i\delta_{nm}$,
connected to the electric field as $gL_n=E_n$.
We work in a compact formulation, where $\theta_n$ becomes an angular variable, and $L_n$ can adopt integer eigenvalues.
Gauss' law appears as the additional constraint
\beq
L_n-L_{n-1}=\phi_n^{\dagger}\phi_n-\frac{1}{2}\left [1-(-1)^n\right].
\label{eq:Gaussferm}
\eeq
Instead of the Hamiltonian~\eqref{eq:Hferm}, it is usual to work with the dimensionless operator
$W=\frac{2}{a g^2}H$, and to define parameters 
$x=\frac{1}{g^2 a^2}$ and $\mu=\frac{2 m}{g^2 a}$.
In this picture, fermions on even and odd sites respectively correspond to the upper and lower components of the spinor representing the fermionic field in the continuum.
Gauss' law~\eqref{eq:Gaussferm} determines the electric field up to a constant $\alpha$ which can be added to $L_n$ and represents the background electric field.

Using a Jordan-Wigner transformation, $\phi_n=\prod_{k<n}(i\sigma_k^{z})\sigma_n^{-}$, 
where $\sigma^{\pm}=\frac{1}{2}(\sigma^{x}\pm i\sigma^{y})$,
model~\eqref{eq:Hferm} can be mapped to a spin Hamiltonian \cite{Banks:1975gq},
\begin{equation}
H= x\sum_{n=0}^{N-2} \left [ \sigma_n^+\sigma_{n+1}^- +  \sigma_n^-\sigma_{n+1}^+ \right ]+\frac{\mu}{2}\sum_{n=0}^{N-1} \left [ 1 + (-1)^n \sigma_n^z \right ]
+\sum_{n=0}^{N-2} \left (L_n+\alpha\right )^2.
\label{eq:Hspin}
\end{equation}
Gauss' law reads then $L_n-L_{n-1}=\frac{1}{2}\left [ \sigma_n^z + (-1)^n \right ]$, and can be used to eliminate the gauge degrees of freedom,
leaving the Hamiltonian \cite{hamer97free}
\begin{align}
H=& x\sum_{n=0}^{N-2} \left [ \sigma_n^+\sigma_{n+1}^- +  \sigma_n^-\sigma_{n+1}^+ \right ]+\frac{\mu}{2}\sum_{n=0}^{N-1} \left [ 1 + (-1)^n \sigma_n^z \right ]
\nonumber \\
&+\sum_{n=0}^{N-2} \left [ \ell +\frac{1}{2}\sum_{k=0}^n ((-1)^k+\sigma_k^{z})\right ]^2,
\label{eq:H}
\end{align}
where $\ell$ is the boundary electric field, on the link to the left of site $0$, which can describe the background field.  

 A useful basis for this problem is then
\beq
|\ell\rangle \otimes | i_0 i_1 \ldots i_{N-2} i_{N-1} \rangle,
\eeq 
with $i_m=\{0,\, 1\}$ labeling the $\pm1$ eigenstates of $\sigma_m^z$ (on site $m$).
An even site in spin state $|1\rangle$ corresponds to the presence of a fermion, 
while an odd site in state $|0\rangle$ is an antifermion  at the corresponding position.
The integer valued $\ell=\ldots,-1,0,1,\ldots$ is the only gauge degree of freedom left,
but with OBC it is non-dynamical, as the Hamiltonian cannot connect states with different values of $\ell$.
Here we choose $\ell=0$ and omit it in the following, as we will be concerned with the case of zero background field.
Moreover, we are interested in states with zero total charge ($\sum_n \sigma_n^z=0$ in the spin language), so 
we consider chains with even $N$.



\section{Method}
\label{sec:method}


In this work we use the MPS ansatz to approximate the ground and lowest excited states 
of the Hamiltonian~\eqref{eq:H}.
A MPS for a system of $N$ $d$-dimensional sites has the form
\beq
 |\Psi\rangle =\sum_{i_0,\ldots
  i_{N-1}=1}^d \mathrm{tr}(A_0^{i_0}\ldots A_{N-1}^{i_{N-1}}) |i_0,\ldots i_{N-1}
\rangle,
\eeq
 where $\{|i\rangle\}_{i=0}^{d-1}$ are  individual basis states for each site.
Each $A_k^i$ is a $D$-dimensional matrix and the bond dimension, 
 $D$, determines the number of parameters in the ansatz.
MPS are known to provide good approximations to ground states of local
Hamiltonians in the gapped phase,
but have also been successfully used for more general models.


The MPS approximation to the ground state can be found variationally 
by successively minimizing the energy, $\frac{\langle\Psi|H|\Psi\rangle}{\langle\Psi|\Psi\rangle}$, with 
respect to each tensor $A_k$ until convergence.
Each such optimization reduces to the eigenvalue problem of an effective Hamiltonian
that acts on site $k$ and its adjacent virtual bonds.
The basic algorithm, closely related to the original DMRG formulation \cite{white92dmrg,schollwoeck05dmrg},
was introduced in \cite{verstraete04dmrg}, and is nowadays widely used.

\begin{figure}[floatfix]
\subfigure[Pictorial representation of the MPS, and Hamiltonian and norm contractions]{
 \label{fig:fig1-a}
\psfrag{P}[c][l]{$|\Psi\rangle$}
\psfrag{A}[c][l]{$A_k$}
\psfrag{H}[c][c]{$\langle\Psi|H|\Psi\rangle$}
\psfrag{N}[c][l]{$\langle\Psi|\Psi\rangle$}
  \includegraphics[width=.85\columnwidth]{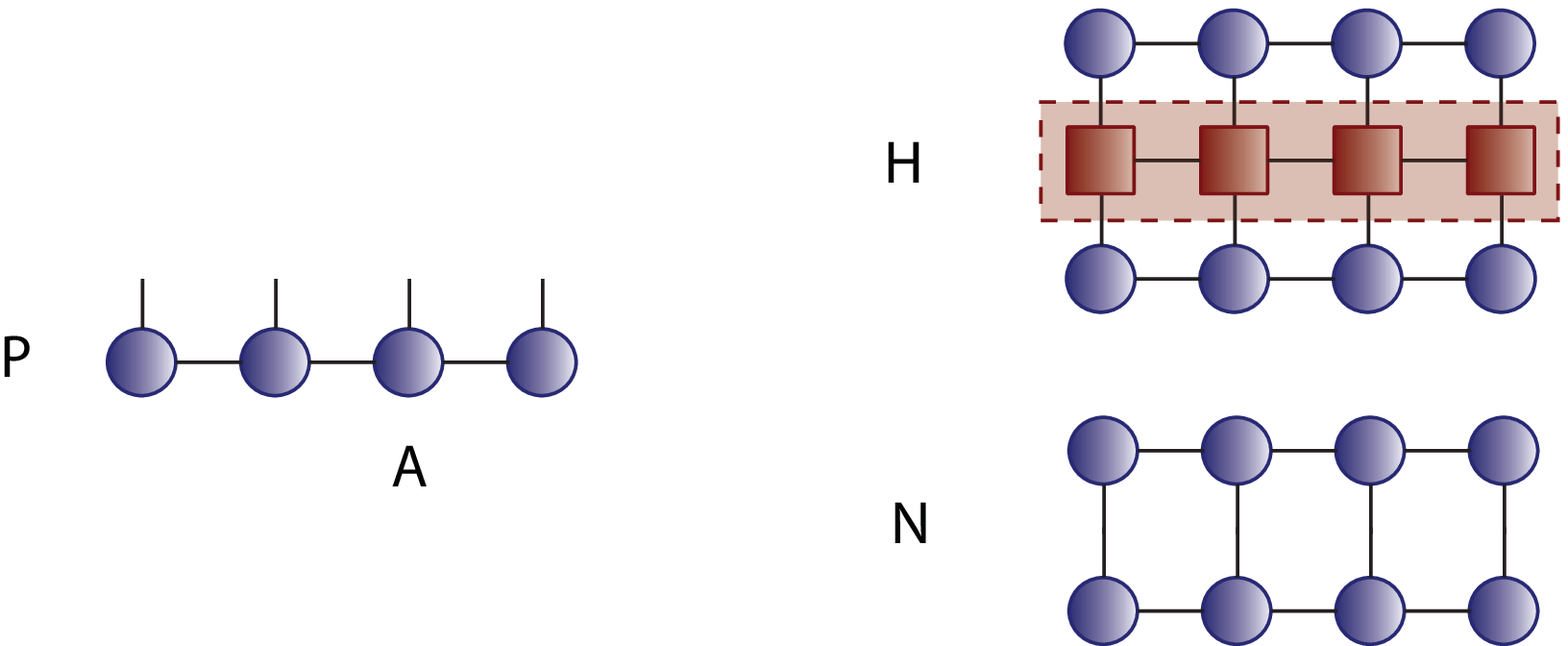}
}
\subfigure[Effective Hamiltonian for site $k$ and excited level $M$.]{
 \label{fig:fig1-b}
\psfrag{H}[c][l]{${\cal{H}}_{\mathrm{eff}}^k$[M]=}
\psfrag{h}[c][l]{${\cal{H}}_{\mathrm{eff}}^k$}
\psfrag{M}[c][l]{$+\sum_{m=0}^{M-1} E_m \times$}
\psfrag{p}[l][l]{$(\Pi_m)_{\mathrm{eff}}^k$}
\psfrag{P}[l][l]{$|\Psi_m\rangle_{\mathrm{eff}}^k$}
  \includegraphics[width=.85\columnwidth]{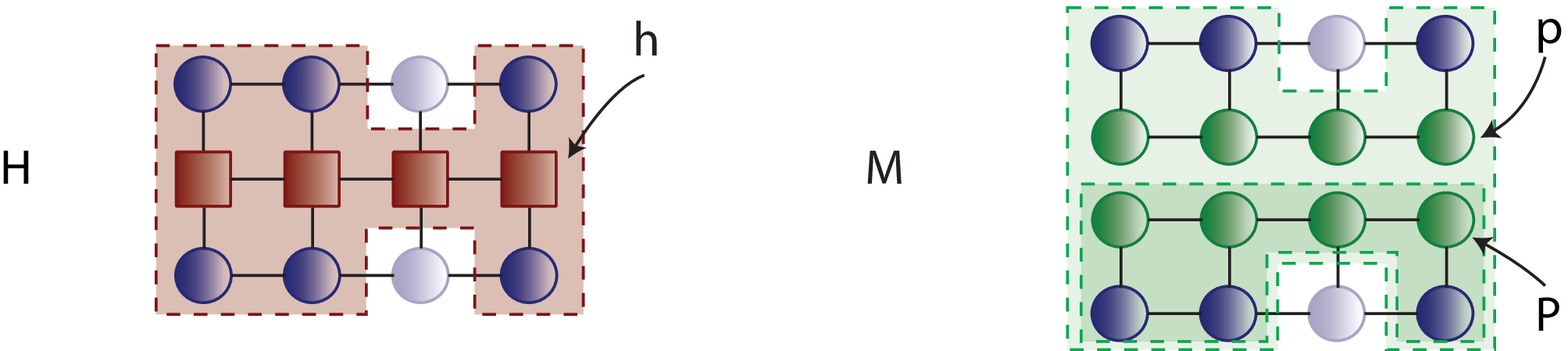}
}
\caption{
Scheme of the algorithm for excited states. 
In \ref{fig:fig1-a} (left) we show the commonly used graphical representation of a MPS (see e.g. \cite{verstraete08algo}). Each circle corresponds to one tensor ($A_k$)
and each of its legs represents one index, with lines that join two tensors representing a contracted index (as in a matrix multiplication).
The open legs correspond to the physical indices on each site.
A particular coefficient in the product basis corresponds to fixing each of the open indices to a value  ($0,\ldots, d-1$).
On the right, we show the representation of some usual contractions. 
We can contract two MPS by joining their open (physical) indices to compute the norm (lower scheme) or insert an operator with a matrix product structure, as the Hamiltonian, to obtain the expectation value of the energy (above).
In \ref{fig:fig1-b} we show the tensor network representation of ${\cal{H}}_k$ in the step of the optimization where site $k$ is computed. 
The term on the left is simply the contraction of the TN for $\langle H\rangle$
except for tensor $A_k$.
Each term in the sum on the right is the TN for the expectation value of one projector ($\langle \Psi|\Psi_m\rangle \langle \Psi_m|\Psi\rangle$)
leaving out the tensor at site $k$.
The sum of all these terms produces the effective Hamiltonian at site $k$.
}
\label{fig:TN}
\end{figure}

In DMRG it is possible to target several of the lowest eigenstates \cite{hallberg06}, or, in the case of a first excited state with a distinct quantum number, to run the ground state search in different sectors.
In the MPS formalism, it is natural to extend the 
 variational method to determine excited states by 
 performing a constrained minimization which imposes orthogonality with respect to the already computed states (ground and lower excited states), as proposed in \cite{porras06spec} and recently extended in \cite{wall12outofeq}.
Here we introduce a simpler variation of the ground state algorithm, which allows us to compute MPS approximations to the lowest energy eigenstates at  lower cost, without using any explicit symmetry. 
This is especially interesting for a finite system with open boundary conditions, where momentum is not a good quantum number. 
It is worth noticing that in the case of translationally invariant (TI) systems, either finite and periodic or infinite,
a very successful approach exists based on the
construction of 
well-defined momentum MPS \cite{haegeman12disp}.

For finite systems, the MPS algorithms for open boundary conditions are numerically more stable and in general more efficient,
although  improved methods have been recently proposed for periodic systems \cite{pippan10pbc,pirvu11ti}.
However, finite-size effects are much larger for OBC and therefore simulation of larger chains may be needed to reach the thermodynamic limit reliably.

After having found the ground state of the system, $|\Psi_0\rangle$,
we can construct the projector onto the orthogonal subspace, $ \Pi_{ 0}=1-|\Psi_0\rangle\langle \Psi_0|$.
The projected Hamiltonian, $\Pi_{0} H \Pi_{ 0}$,
has $|\Psi_0\rangle$ as eigenstate with zero eigenvalue, and 
the first excited state as eigenstate with energy $E_1$.
Given that $E_1<0$, what we can always ensure by adding an appropriate constant to $H$,
the first excitation corresponds then to the state that minimizes the energy of the projected Hamiltonian,
\footnote{Alternatively, we can use a Hamiltonian $H+\Delta |\Psi_0\rangle\langle \Psi_0|$,
with $\Delta>E_1-E_0$.}
\beq
E_1=\min_{|\Psi\rangle}\frac{\langle \Psi|\Pi_{0} H \Pi_{ 0}|\Psi\rangle}{\langle \Psi|\Psi\rangle}=
\frac{\langle \Psi|\left(H-E_0 |\Psi_0\rangle\langle \Psi_0|\right)|\Psi\rangle}{\langle \Psi|\Psi\rangle}.
\eeq
This minimization corresponds to finding the ground state of the effective Hamiltonian 
$H_{\mathrm{eff}}[1]=\Pi_0 H\Pi_0$.
The procedure can be concatenated to find subsequent energy levels, so that, to find the $M$-th excited state, we will 
search for the ground state of the Hamiltonian
\begin{equation}
H_{\mathrm{eff}}[M]=\Pi_{M-1}\ldots\Pi_0 H \Pi_0\ldots \Pi_{M-1}
= H-\sum_{k=0}^{M-1} E_k |\Psi_k\rangle \langle \Psi_k|.
\label{eq:HeffM}
\end{equation}

\begin{figure}[htbp]
\begin{algorithmic}
\REQUIRE{Hamiltonian for an $N$-site chain, $H$; maximum bond dimension, $D$; tolerance, $\epsilon$}
\ENSURE{MPS approximation to the ground state, $|\Psi \{A_k\}\rangle_D$, with energy $E$}
\STATE $\{A_0,\ldots A_{N-1}\} \leftarrow$ initial guess
\STATE $E \leftarrow \frac{ \langle \Psi | H |\Psi\rangle}{ \langle \Psi |\Psi\rangle}$
\STATE $\delta E \leftarrow 1$
\STATE $sweeping\_direction \leftarrow $ right
\WHILE{$\delta E >\epsilon$}
\STATE $k \leftarrow$ first site for $sweeping\_direction$ 
\WHILE{$0\leq k\leq N-1$}
\STATE compute matrices ${\cal{H}}_{\mathrm{eff}}^k$, ${\cal{N}}_{\mathrm{eff}}^k$  \COMMENT{contraction without tensor $A_k$ (Fig. \ref{fig:TN})}
\STATE solve generalized eigenvalue problem ${\cal{H}}_{\mathrm{eff}}^{k} A=\lambda_{\mathrm{min}} {\cal{N}}_{\mathrm{eff}}^{k} A$
\STATE $A_k \leftarrow A$, $E_k \leftarrow \lambda_{\mathrm{min}}$
\STATE $k \leftarrow$ next site according to $sweeping\_direction$
\ENDWHILE
\STATE $\delta E \leftarrow \left |\frac{E-E_k}{E}\right |$
\STATE $E\leftarrow E_k$
\STATE flip $sweeping\_direction$ \COMMENT{left $\leftrightarrow$ right}
\ENDWHILE
\end{algorithmic}
\captionof{figure}{Schematic MPS variational algorithm for the ground state search. An efficient implementation
requires imposing a canonical form of the MPS after each update, and reusing
temporary calculations, among other optimizations (see \cite{verstraete08algo,schollwoeck11age}).}
\label{fig:mps-code}
\end{figure}

 Each of these ground state searches can be solved by applying the standard variational MPS algorithm 
to the corresponding effective Hamiltonian~\eqref{eq:HeffM}, which can be constructed from the set of all previously 
computed MPS levels $\{|\Psi_k\rangle\}$ (see Fig.~\ref{fig:mps-code}).
The expression to be minimized at each step of the MPS iteration is then
$\min_{A_k}\frac{A_k^*{\cal{H}}_kA_k}{A_k^*{\cal{N}}_kA_k}$, 
where ${\cal{H}}_{k}$ and ${\cal{N}}_{k}$ are the effective Hamiltonian and norm matrix acting on site $k$ and its 
two adjacent virtual bonds,
so that  each tensor can be found by solving a generalized eigenvalue problem ${\cal{H}}_{k} A_k=\lambda {\cal{N}}_{k} A_k$.
As illustrated in figure~\ref{fig:TN}, 
${\cal{H}}_{k}$ and ${\cal{N}}_{k}$ are computed by contracting all tensors but $A_k$, and
the computational cost can be optimized, as in the original algorithm,  by storing and reusing the partial contractions
that compose these effective matrices.
In general this is most easily done when all the terms in the problem Hamiltonian are expressed as a matrix product operator 
(MPO) \cite{pirvu10mpo}. 
In this particular case it is more convenient to keep separate temporary terms for
each of the contractions $\langle\Psi_k|\Psi\rangle$, and to reconstruct from them 
the effective projectors on every site, before constructing the effective Hamiltonian.
In this way, the number of operations required for each such term scales as $d D^3$, 
without increasing the leading cost of the algorithm.
Due to the larger number of terms that need to be kept, however, 
the cost of finding the $M$-th level once all the lower ones are known will be  
about $M$ times higher than that of the original ground state search~\footnote{This refers only to the leading computational cost, as in some cases convergence of a particular excited states may result slower due to small gaps.},
and thus the total cost for computing up to the $M$-th level will scale as $M^2 D^3$. 

\begin{figure}[floatfix]
\begin{minipage}[b]{.49\columnwidth}
\subfigure[$m/g=0$]{
\includegraphics[height=.9\columnwidth]{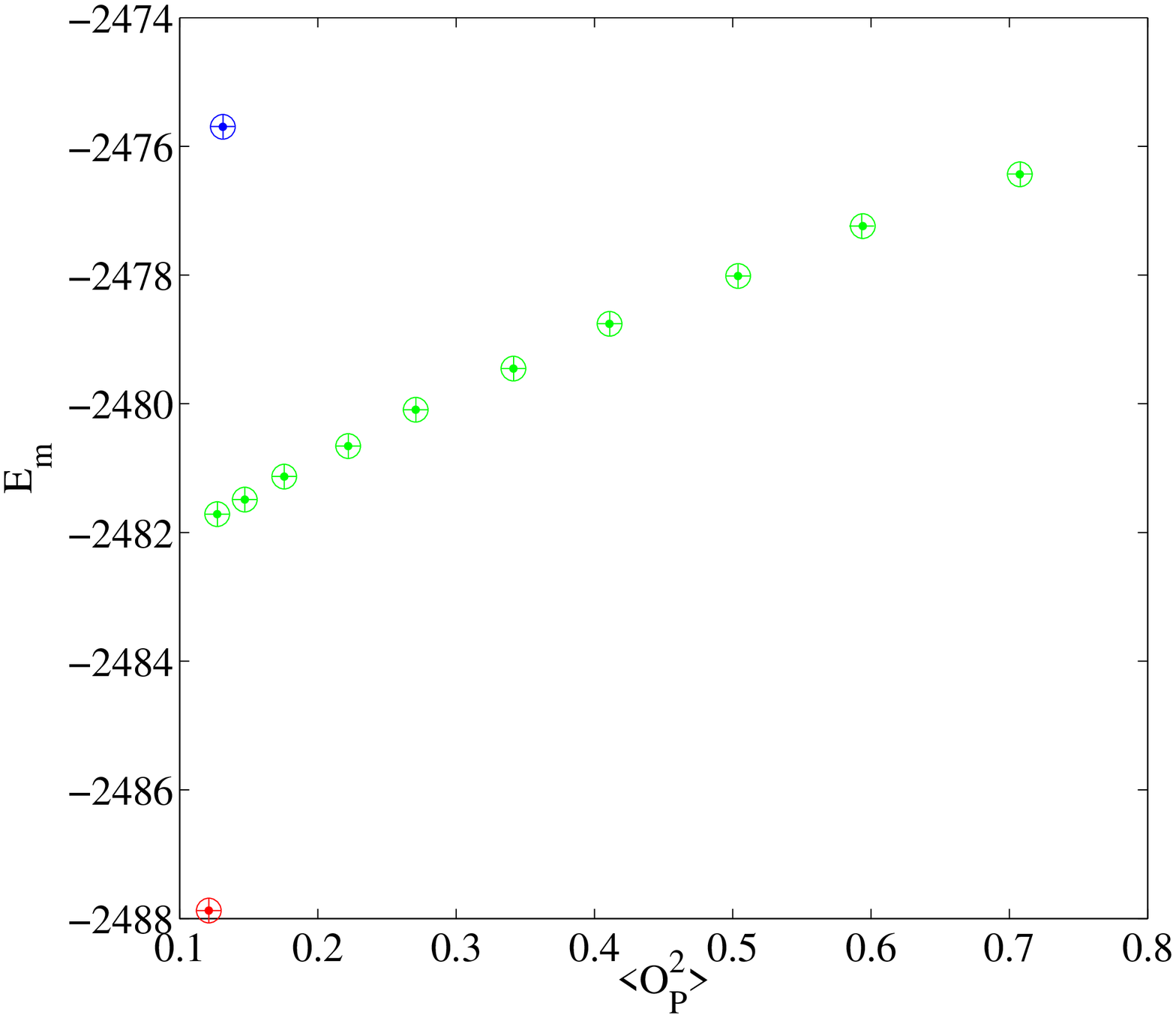}
}
\end{minipage}
\begin{minipage}[b]{.49\columnwidth}
\subfigure[$m/g=0.125$]{
\includegraphics[height=.9\columnwidth]{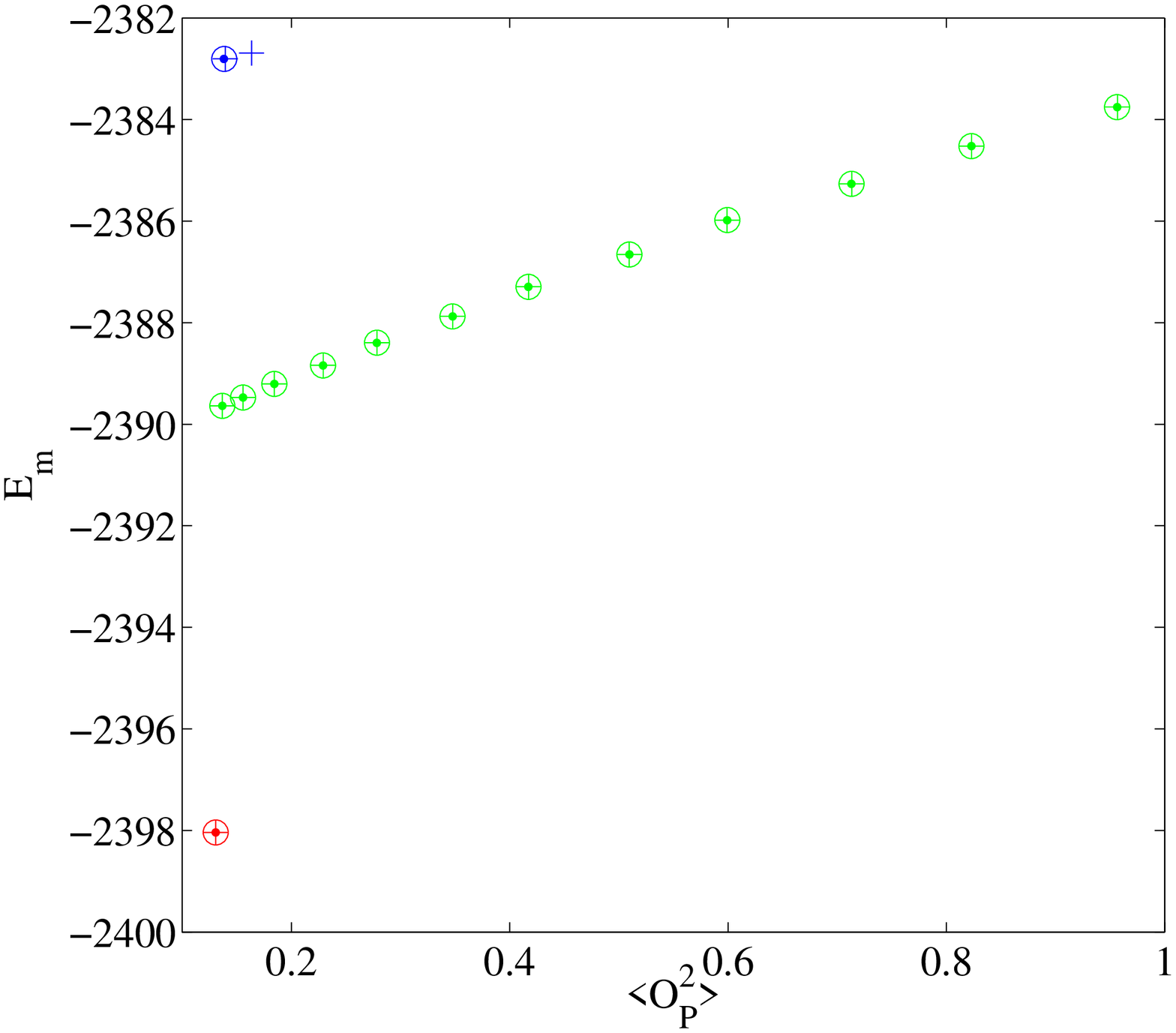}
}
\end{minipage}
\caption{Dispersion relation (energy vs. $\langle O_P^2\rangle$) obtained with MPS for $x=25$, $N=160$
and $m/g=0$ (left) and $m/g=0.125$ (right). 
Shown are the states reconstructed with bond dimensions $D=40$ (crosses), $80$ (circles) and $100$ (dots)
 (being on top of each other in the graph) until we have reached a scalar
candidate.
The appearance of the scalar is detected by the phase of the expectation value $\langle S_R\rangle$,
with the scalar (indicated in blue in the plots) being the first state 
with $\varphi \approx 0$ above the ground state (red) while states with
$|\varphi|\approx \pi$
belong to the vector branch (green).}
\label{fig:dispersion}
\end{figure}

The masses of the particles in the theory are given by the energies
of the zero momentum excitations.
In finite systems with OBC momentum is difficult to identify 
and we need to find 
the excitations that will correspond to the lowest 
momentum in the continuum limit.
In dynamical DMRG \cite{jeckelmann02}
 momentum dependent quantities were extracted from finite DMRG calculations
 with open boundaries
using quasimomentum states defined from the eigenstates of a free particle in a box.
Here we use a different approach, based on the continuum momentum operator
 for the fermion field,  
$\hat{P}=\int dx \Psi^{\dagger}(x) i \partial_x \Psi(x)$.
Its discretization yields, in the spin representation, and after rescaling by a factor $\frac{2}{a g^2}$, the operator
\beq
\hat{O}_P=-i x \sum_n [\sigma_n^- \sigma_{n+1}^z
      \sigma_{n+2}^+ - h.c.].
      \label{eq:momspin}
\eeq
The expectation values of $\hat{O}_P^2$ can be used to assign  a pseudo-momentum to the spectral levels, in order to
reconstruct the dispersion relation (Fig.~\ref{fig:dispersion}).

The model gives rise to stable particles, the lowest ones being, in the case of no background field, 
the \emph{vector} and \emph{scalar} states.
In the continuum model, these particles are distinguished by well-defined 
parity and charge conjugation quantum numbers \cite{coleman76}, with the scalar living in the 
same sector as the ground state.
On a staggered lattice with PBC it is possible to exploit the corresponding lattice symmetries
to construct two orthogonal subspaces, one of them containing the vector, and the other the ground state 
and the scalar \cite{banks76sce}.
For a chain with OBC the number of surviving symmetries is even lower, with translational 
invariance lost.
In practice this means that to calculate the scalar mass, we need to identify first the momentum
excitations of the vector, which appear at lower energy than the scalar.
In \cite{hamer97free} this was achieved by starting from the strong coupling
limit ($x\to0$), where vector and scalar states are known exactly, and
smoothly changing the $x$ parameter while keeping a label on the scalar state. 
Instead, we use the expectation value of the spin transformation,
$S_R=\otimes_{k=0}^{N-1} \sigma_{2k-1}^{x} T^{(1)}$,
where $T^{(1)}$ is the (cyclic) translation by one spin site~\footnote{The choice of a cyclic translation ensures unitarity of the operator, and is irrelevant in the thermodynamic limit.}.
In a system with PBC this operator basically describes the action of charge conjugation on the spins, with the translation exchanging the fermionic and antifermionic character of sites, and the $\sigma_x$ rotation accounting for the different meaning of a spin \emph{up} on an even 
(fermion) site (empty) and on an odd one (occupied).
Charge conjugation is a good quantum number and distinguishes the vector state ($C=-1$) from the sector containing the scalar and the 
ground state ($C=+1$), even for finite systems.
On the staggered lattice with OBC this is no longer true, and $S_R$ does not commute with the Hamiltonian.
However the phase of $S_R$ keeps memory of this distinction, and allows us to tag the levels accordingly, with the ground state and the scalar branch of the dispersion relation having
phase $\varphi (\langle S_R\rangle)\approx 0$ and the vector branch $\varphi(\langle S_R\rangle) \approx\pi$,
as illustrated in figure~Ê\ref{fig:dispersion}.

\section{Results}
\label{sec:results}

We have applied the methods described in section~\ref{sec:method} to the Hamiltonian~\eqref{eq:H}
to determine the ground state, and the vector and scalar mass gaps for 
various 
 values of the fermion mass. 
In order to benchmark our results with existing data, we have studied different masses $m/g=0$, $0.125$,
$0.25$ and $0.5$, for which reference values can be found in the literature.

Our goal is
to compute, for each of these masses, the ground state energy density, and the vector and scalar mass gaps in
the 
thermodynamic limit ($N\to \infty$) for different finite values of $x$, 
and to extrapolate them to the continuum limit,
corresponding to the lattice spacing $a\to 0$, or $x\to\infty$.
We have used values of $x\in[5,600]$.
In order to extract the thermodynamic limit for each case, we need to simulate different system sizes, $N$,
and apply finite-size scaling.
The system sizes, $N$, cannot be chosen
 independently of the value of $x$, as the same number of sites $N$ corresponds to different
physical volumes $L_{\mathrm{phys}}=N a$  for different values of $x\propto 1/a^2$.
From numerical simulations, we estimated that  $N\geq 20 \sqrt{x} $ 
ensures small enough finite volume effects.
We thus study for each value of $x$ 
several different system sizes,
large enough for the
condition above to be satisfied (for instance, for $x=5$ we take $N\in[50,82]$ while for $x=600$,
$N\in[540,834]$). 

We have run the MPS algorithms described in the previous section 
for each set of parameters, $(m/g,x,N)$,  using bond dimensions $D\in[20,140]$,
to find the ground state and excited levels  
at least until a candidate scalar state is found. 
We are interested in the subspace of null total charge, which corresponds to $\sum S_z=0$ and can 
easily be imposed by adding a penalty term to the Hamiltonian.
 For every level, the MPS iteration stops when the relative change in the energy after one full sweep over the chain 
is below a certain tolerance, $\epsilon$.
This value has to be small to ensure a good precision after the extrapolations, but a smaller tolerance
translates in more difficult convergence. Therefore we fixed $\epsilon=10^{-12}$ for the ground state and the
vector calculations, and $\epsilon=10^{-7}$ for the longer computations required by the
scalar states.

\begin{figure}[floatfix]
\begin{minipage}[b]{.35\columnwidth}
\subfigure{
\includegraphics[height=.8\columnwidth]{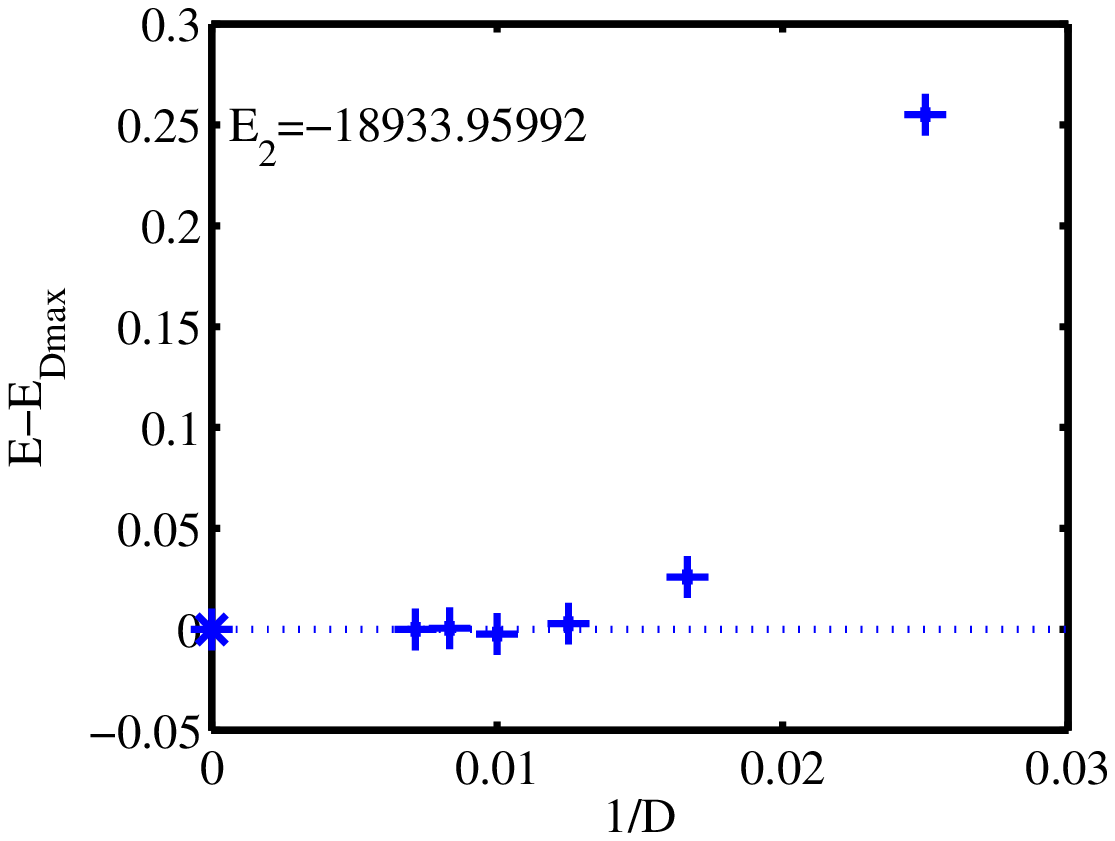}
}
\\
\subfigure{
\includegraphics[height=.8\columnwidth]{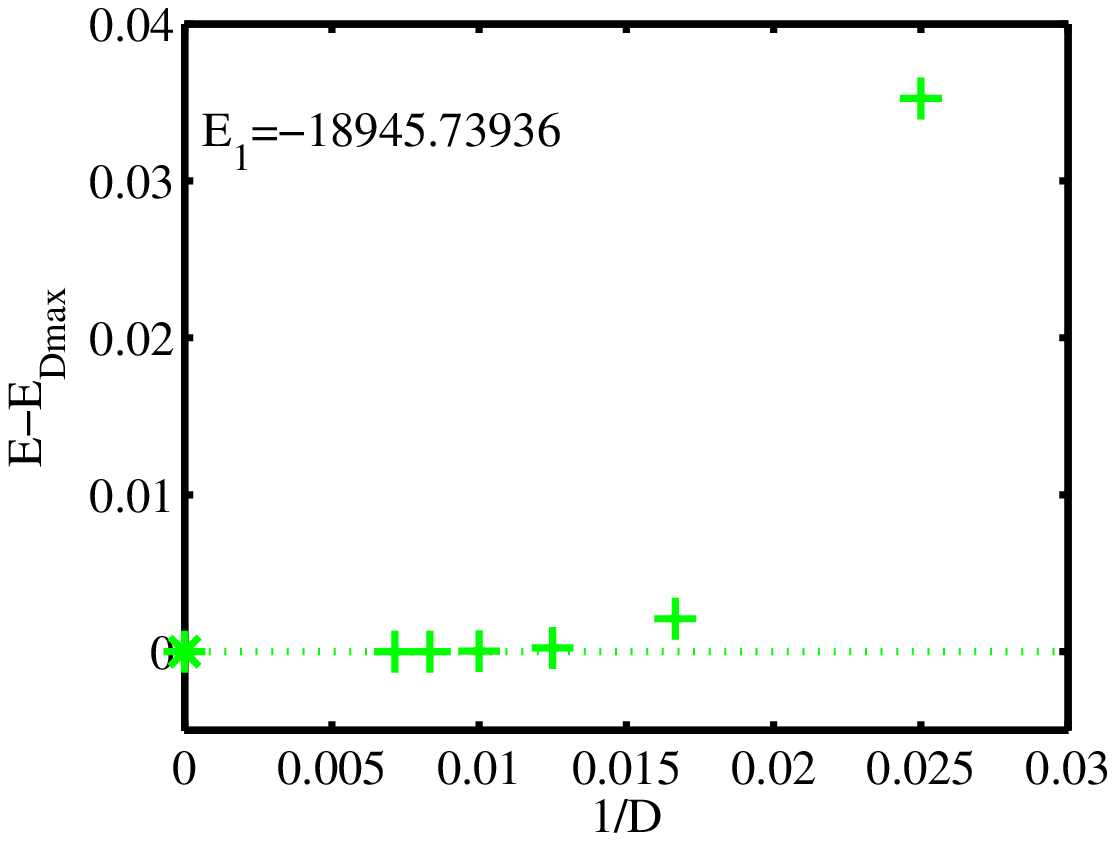}
}
\\
\subfigure[$m/g=0$]{
\includegraphics[height=.8\columnwidth]{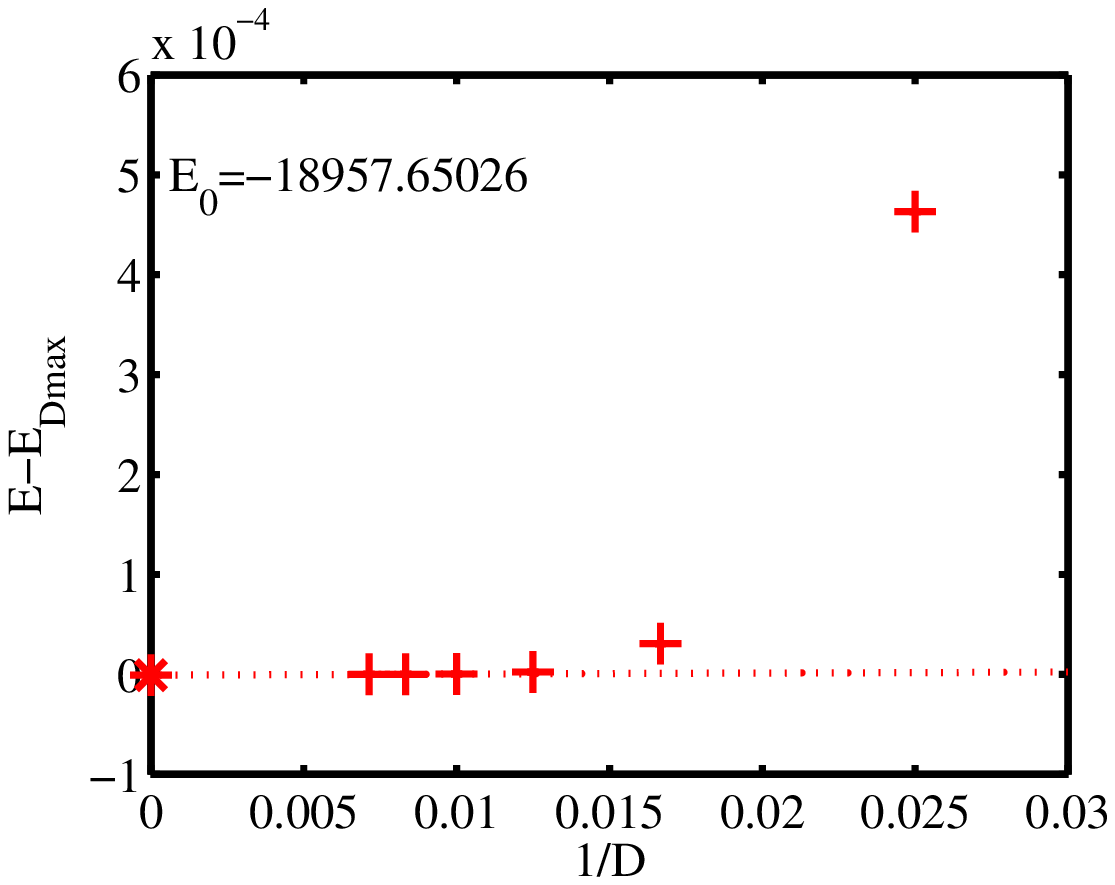}
}
\end{minipage}
\hspace{.1\columnwidth}
\begin{minipage}[b]{.35\columnwidth}
\subfigure{
\includegraphics[height=.8\columnwidth]{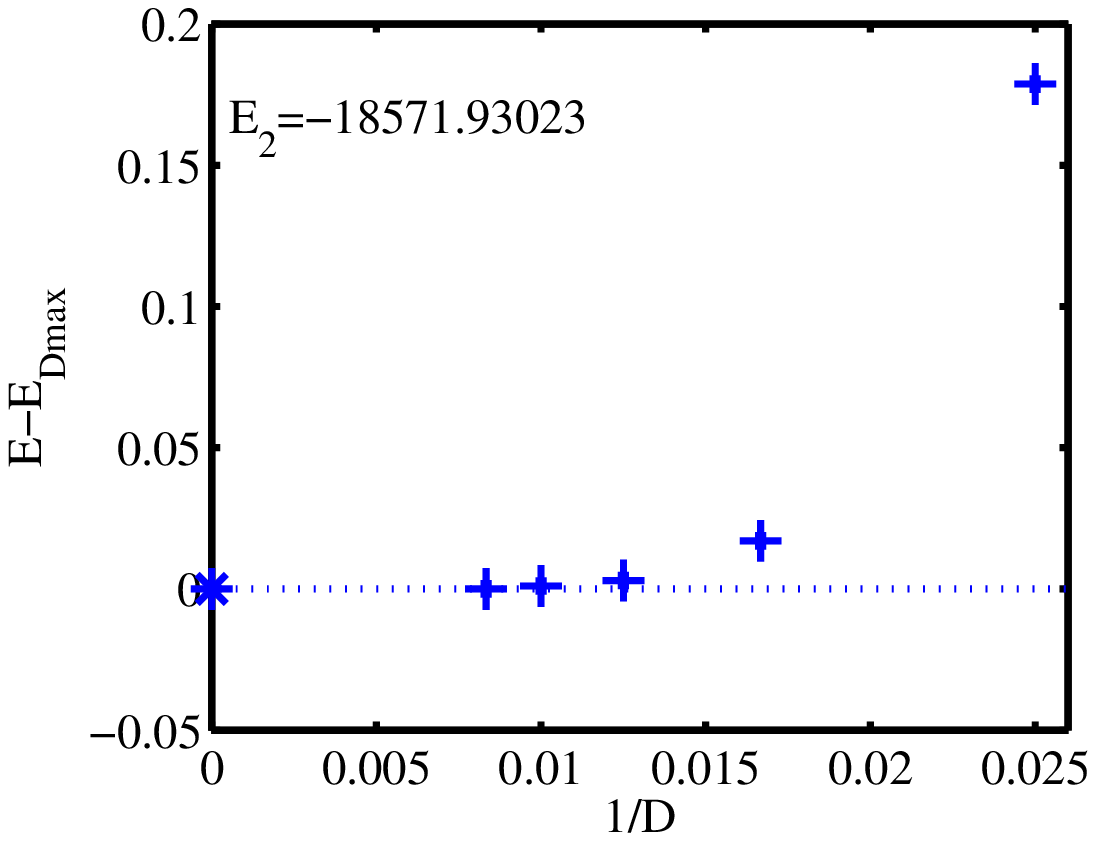}
}
\\
\subfigure{
\includegraphics[height=.8\columnwidth]{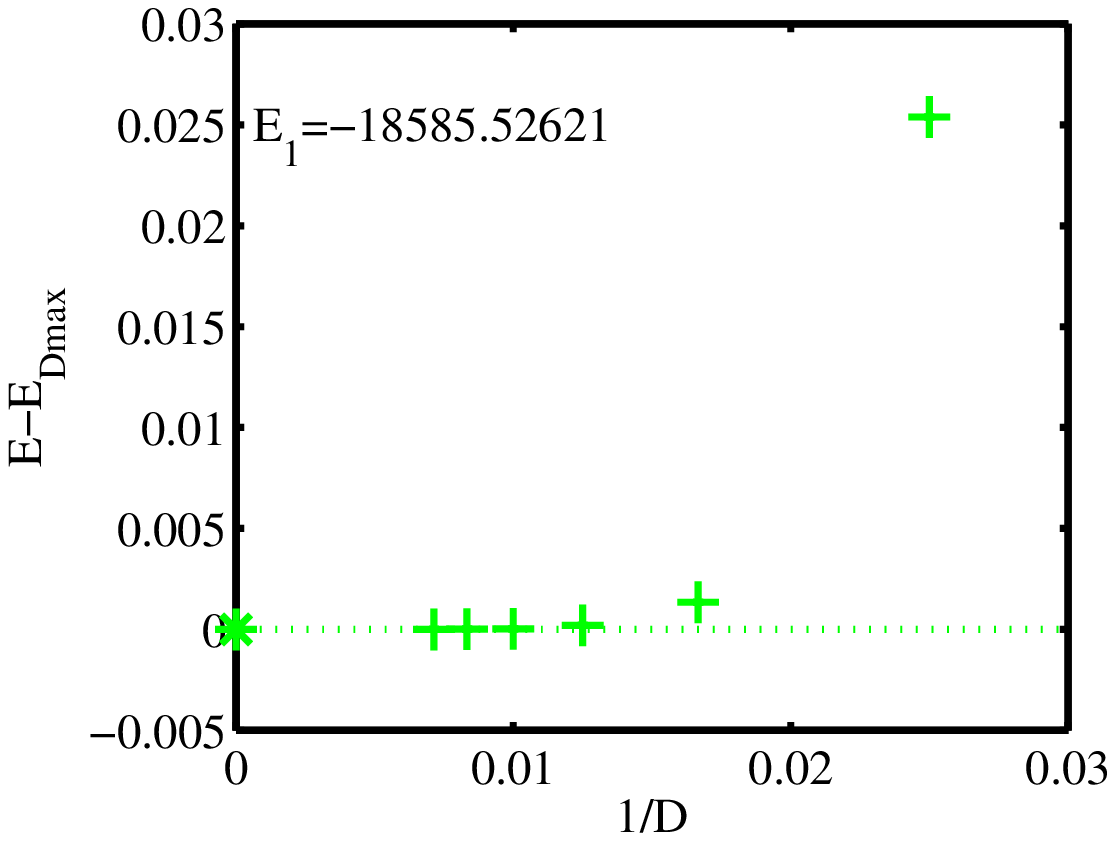}
}
\\
\subfigure[$m/g=0.125$]{
\includegraphics[height=.8\columnwidth]{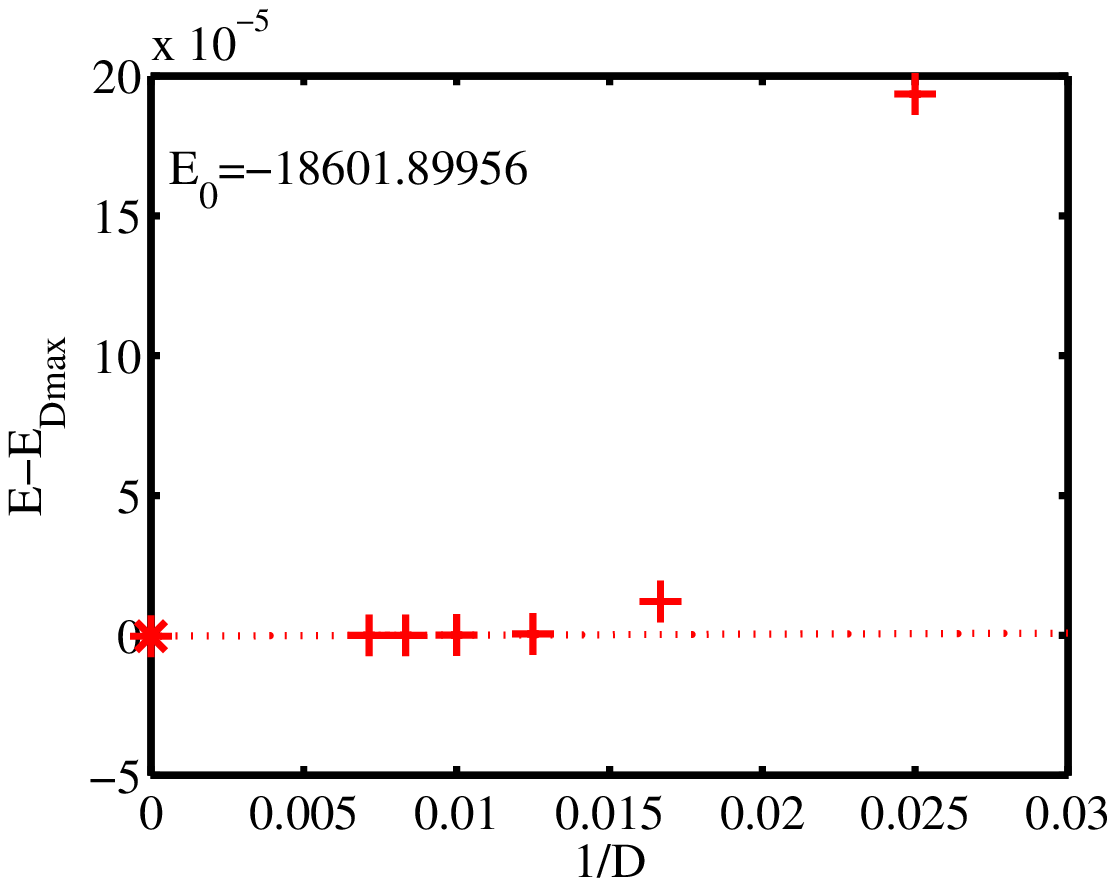}
}
\end{minipage}
\caption{
Convergence of the energy levels with the bond dimension, $D$, for $x=100$ and $N=300$. In
the left column, we show the results of the scalar (uppermost plot, blue),
vector (middle plot, green) and ground state (lowest plot, red) for the massless case, $m/g=0$,
while the right column contains the corresponding results for the case $m/g=0.125$.
Plotted is the difference between the computed energy at a certain bond dimension and the one for the maximum $D_{\mathrm{max}}$. 
The error bars (sometimes smaller than the marker size) indicate the convergence criterion of the MPS algorithm, 
($\epsilon=10^{-7}$ for the scalar and $10^{-12}$ for the rest).
Dashed lines show the extrapolation in $1/D$, with the final value displayed as a star (numerical value inside each panel). 
}
\label{fig:Dscaling}
\end{figure}

\begin{figure}[floatfix]
\begin{minipage}[b]{.4\columnwidth}
\subfigure{
\includegraphics[height=.8\columnwidth]{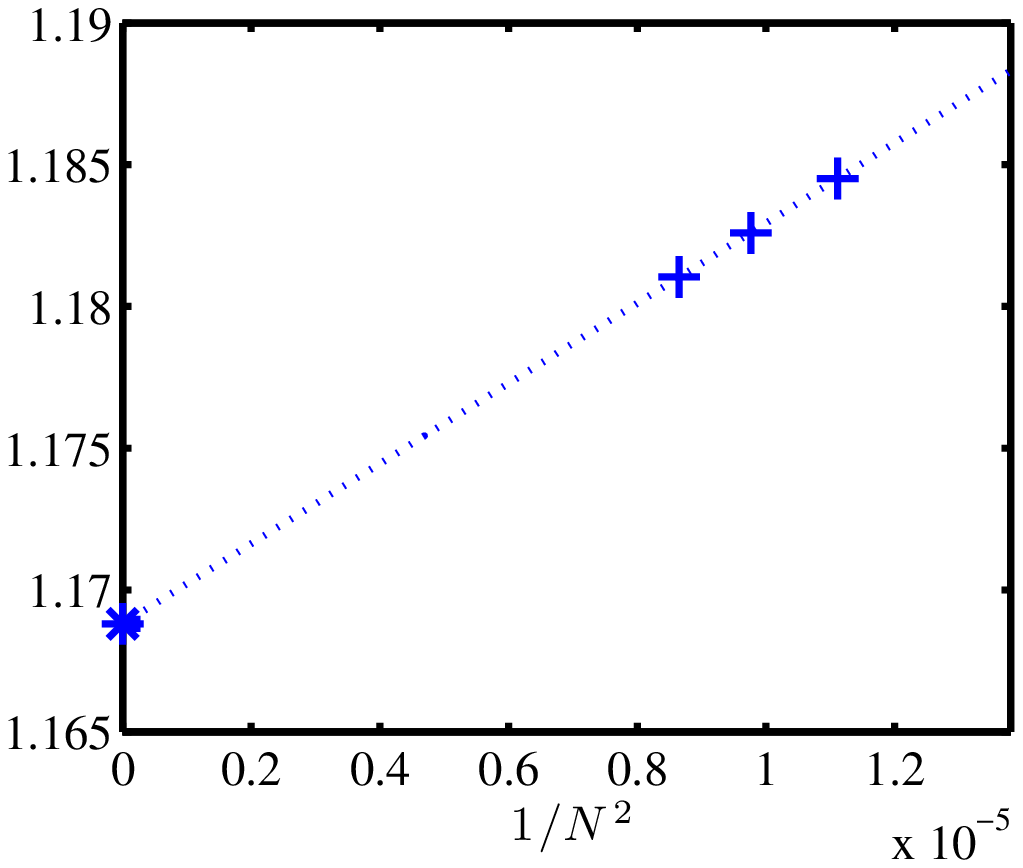}
}
\\
\subfigure{
\includegraphics[height=.8\columnwidth]{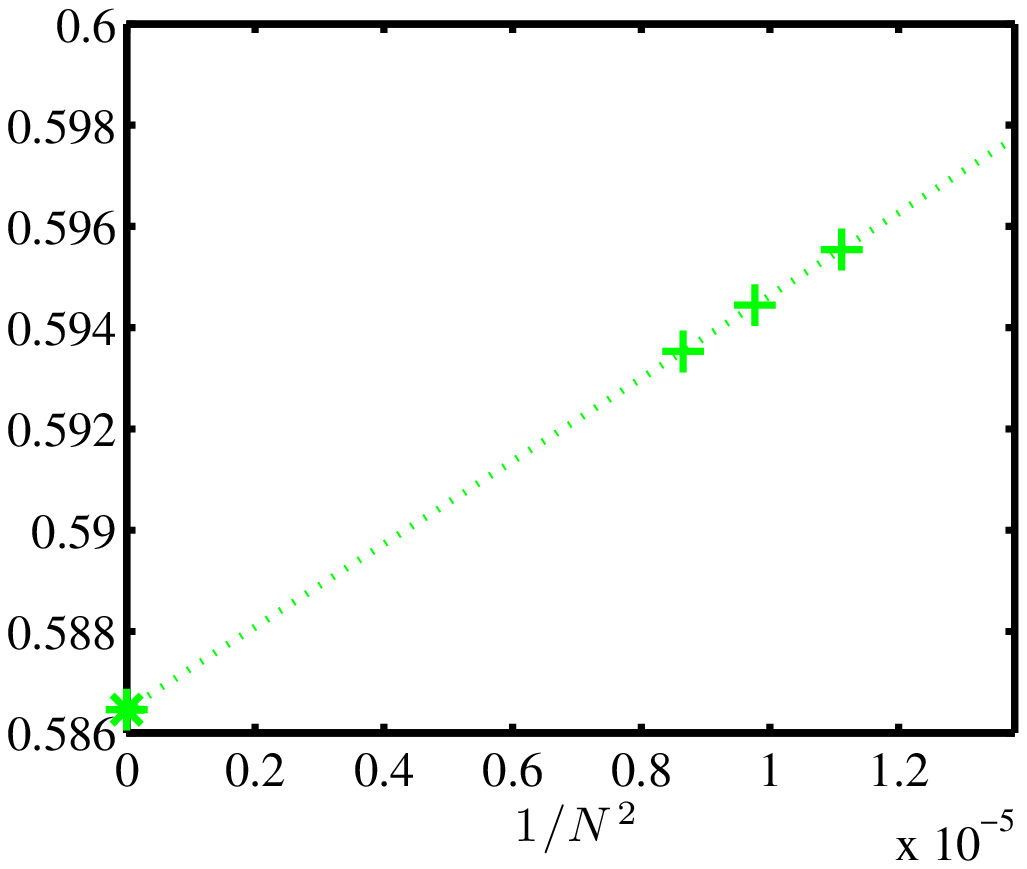}
}
\\
\subfigure[$m/g=0$]{
\includegraphics[height=.8\columnwidth]{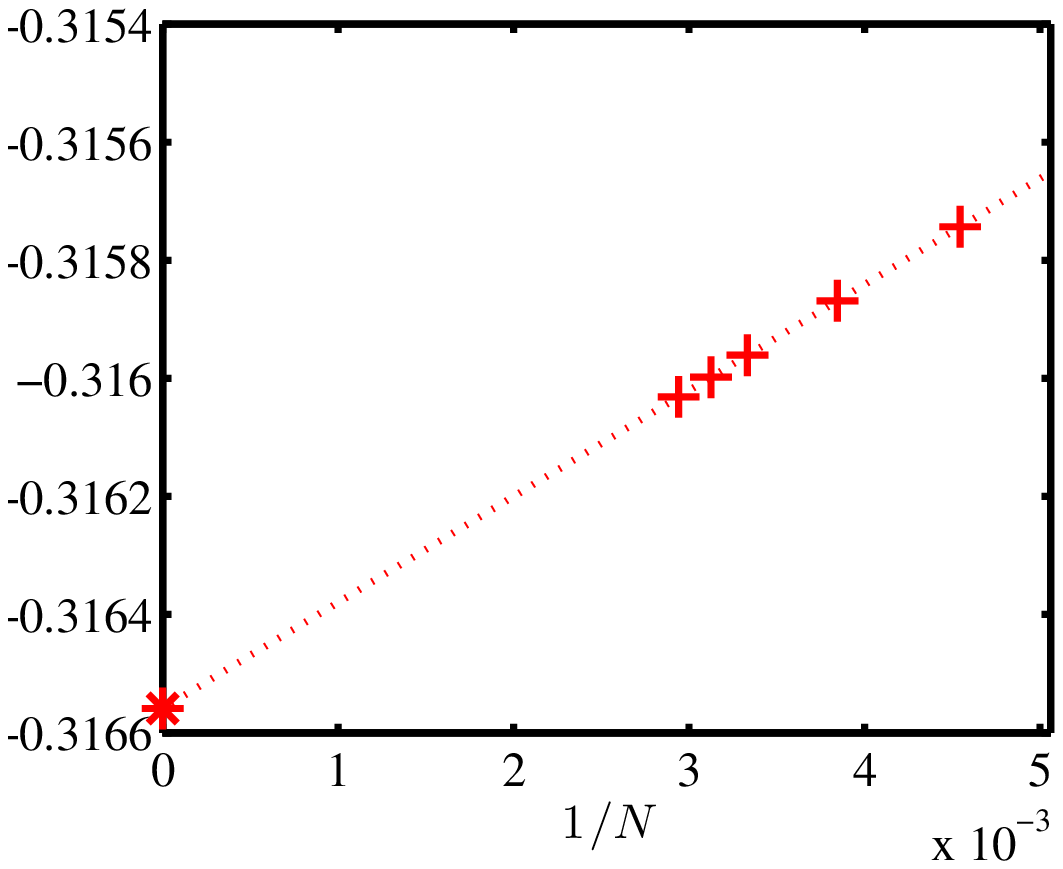}
}
\end{minipage}
\hspace{.1\columnwidth}
\begin{minipage}[b]{.4\columnwidth}
\subfigure{
\includegraphics[height=.8\columnwidth]{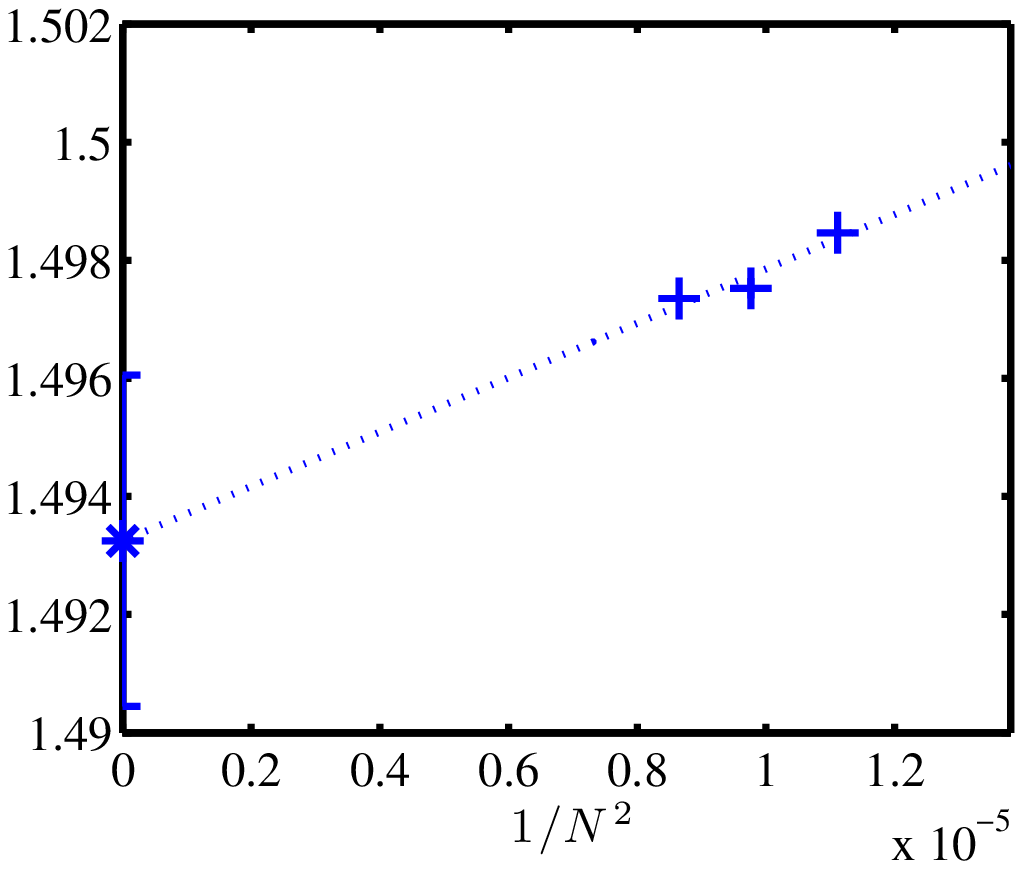}
}
\\
\subfigure{
\includegraphics[height=.8\columnwidth]{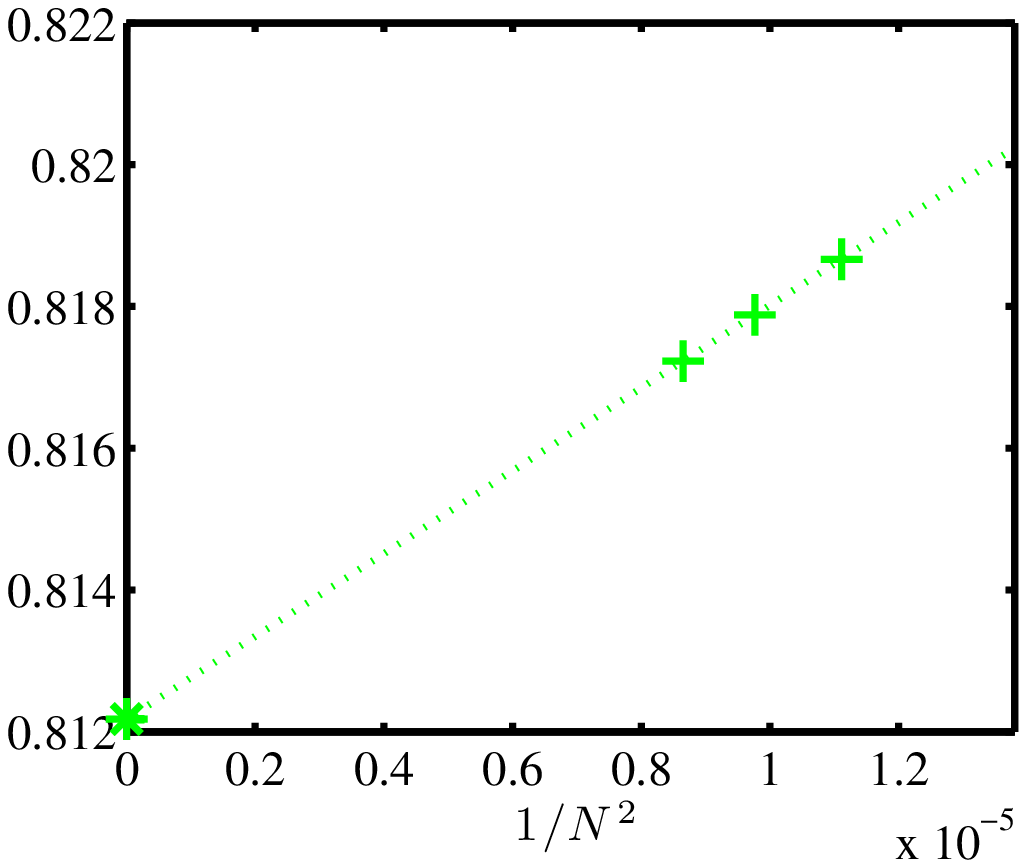}
}
\\
\subfigure[$m/g=0.125$]{
\includegraphics[height=.8\columnwidth]{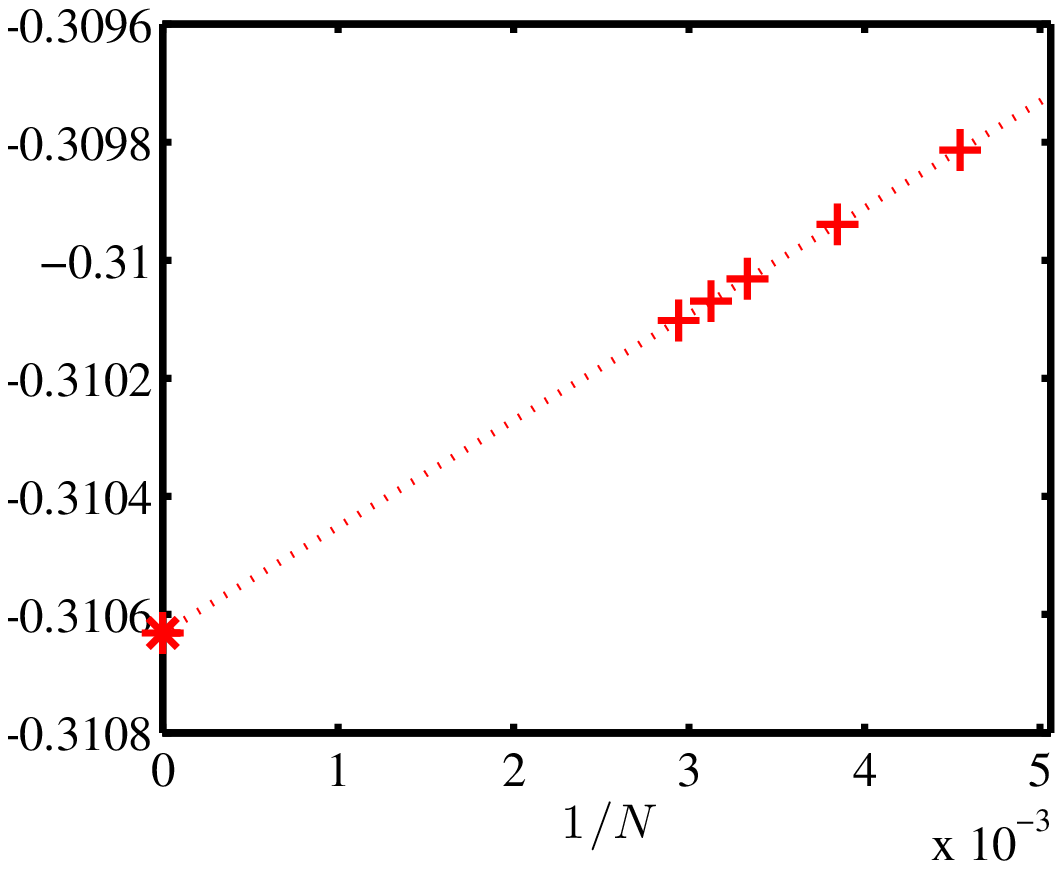}
}
\end{minipage}
\caption{Finite size scaling for $x=100$ and mass $m/g=0$ (left column) and $m/g=0.125$ (right column). The two uppermost rows  show as a function of $1/N^2$
 the values obtained for finite systems for the scalar (blue) and vector (green) mass gaps, $\frac{E_{2(1)}-E_0}{2 \sqrt{x}}$.
 The dashed line corresponds to the fit~\eqref{eq:fitScaling2}, with the resulting value indicated on the axis.
 The lowest row corresponds to the ground state energy density, $\frac{E_0}{2 x N}$, as a function of $1/N$,
 and is fitted according to~\eqref{eq:fitScaling1}.
 Notice that the lattice sizes are already close to the thermodynamic limit.
 The error bars are extracted from the extrapolation in $D$, shown in
figure~\ref{fig:Dscaling}.}
\label{fig:finitesize}
\end{figure}

For every set $(m/g,\,x,\,N)$ and for each of the levels we are interested in, we extrapolate our results to $1/D\to 0$,
as illustrated in figure~\ref{fig:Dscaling}.
In the case of 
 the ground state and the vector state, almost all the bond dimensions are converged, while
for the scalar mass gap, depending on a larger number of intermediate excited states, some of the larger systems are only computed with 
smaller bond dimension. 
If the bond dimension reached is large enough, we extrapolate linearly in $1/D$.
Otherwise we take the value corresponding to the 
largest $D$ as our estimation for the energy, and the error as the difference
between this value and the one for the immediately smaller $D$.

The results of the extrapolation described above provide 
accurate estimators for the energy levels
for various finite chains. 
We then proceed to scale these results with 
the finite system size
 to extract the ground state energy density, $E_0/(2 x N)$, and the mass gaps, $(E_{1(2)}-E_0)/(2\sqrt{x})$
in the thermodynamic limit for each pair of parameters $(m/g,x)$.
Finite-size corrections to the ground state energy density are linear in $1/N$, while for the energy gaps, the leading 
corrections arise from a kinetic energy term $O(\pi/N^2)$  \cite{hamer97free}.
Hence the bulk limit is extracted by fitting the ground state energy density as 
\beq
\frac{E_0}{2 x N}\approx \omega_0+\frac{\alpha}{N}+\frac{\beta}{N^2},
\label{eq:fitScaling1}
\eeq
and the energy gaps as
\beq
\frac{E_{1(2)}-E_0}{2 \sqrt{x}}\approx \omega_{1(2)}+\frac{\alpha_{1(2)}}{N^2}+O\left(\frac{1}{N^3}\right).
\label{eq:fitScaling2}
\eeq
Figure~\ref{fig:finitesize} illustrates this step for $x=100$ and masses $m/g=0$ and $0.125$.

Finally, the values in the continuum limit can be extracted 
by fitting the results for each value of the mass to a polynomial in $\frac{1}{\sqrt{x}}$.
We include only those values of $x$ for which the 
thermodynamic limit could be extracted accurately 
(i.e. the corresponding level for the various system sizes was computed for a large enough $D$). 
The interval $[x_{\mathrm{min}},x_{\mathrm{max}}]$ and the degree of the polynomial 
used for the fit introduce the largest uncertainty in the final
estimators. 
We have thus performed a systematic error analysis 
in which the different 
possible 
fits are taken into account
to estimate the continuum values and their errors from our data.
The detailed method is described in the Appendix \ref{sec:appendix}.

\begin{figure}[floatfix]
\begin{minipage}[b]{.4\columnwidth}
\subfigure[~Ê$m/g=0$, $\frac{M_V}{g}=0.56421(9)$]{ 
\includegraphics[height=.8\columnwidth]{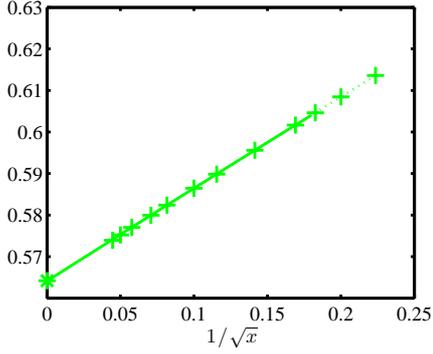}
}
\\
\subfigure[~Ê$m/g=0.125$, $\frac{M_V}{g}=0.53953(5)$]{ 
\includegraphics[height=.8\columnwidth]{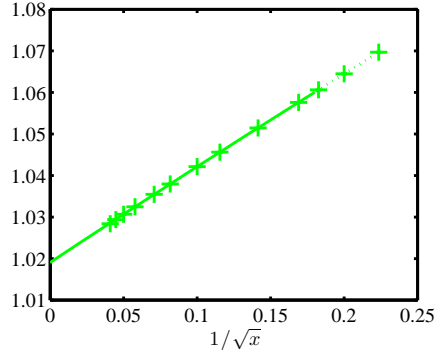}
}
\end{minipage}
\hspace{.1\columnwidth}
\begin{minipage}[b]{.4\columnwidth}
\subfigure[~Ê$m/g=0.25$, $\frac{M_V}{g}=0.51922(5)$]{ 
\includegraphics[height=.8\columnwidth]{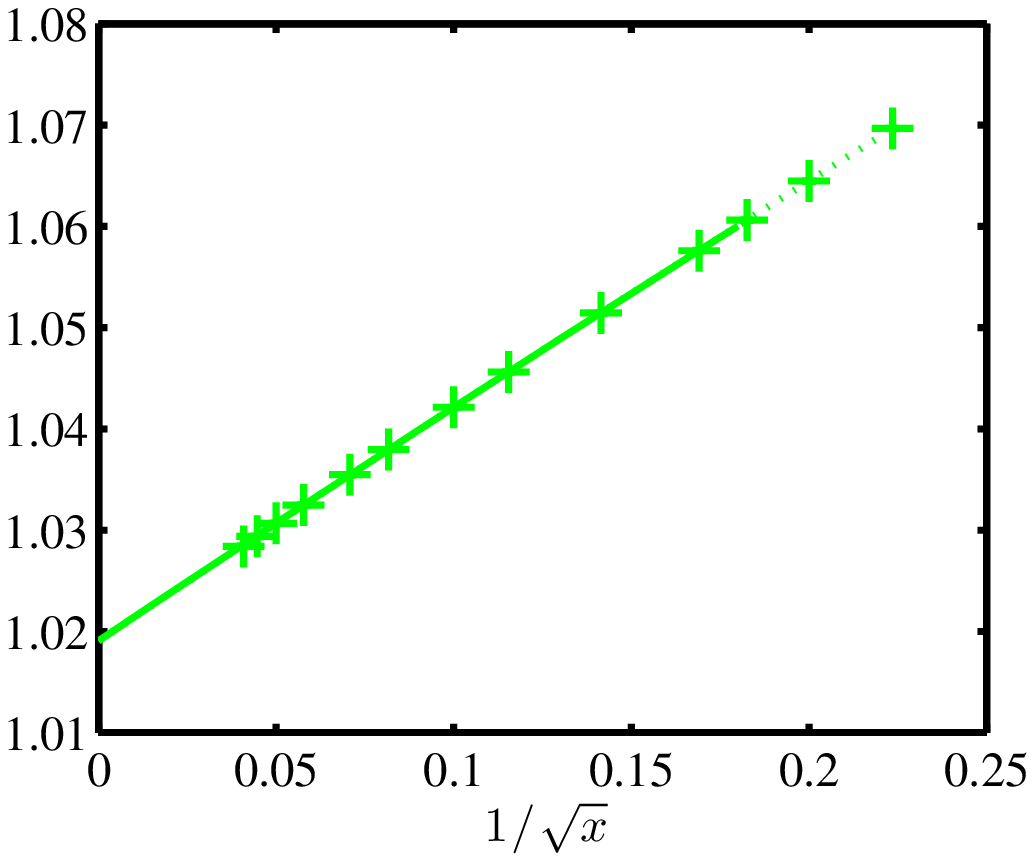}
}
\\
\subfigure[~Ê$m/g=0.5$, $\frac{M_V}{g}=0.48749(3)$]{
\includegraphics[height=.8\columnwidth]{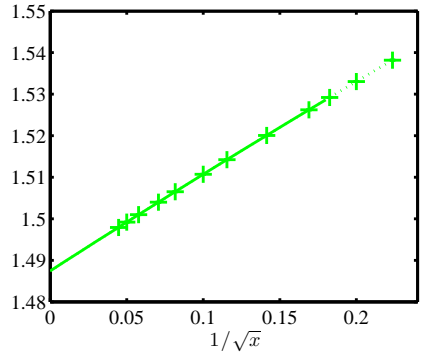}
}
\end{minipage}
\caption{
Energy gap for the vector state, $\omega_1$,
 as a function of $1/\sqrt{x}$.
The uncertainties of the quadratic fits~\eqref{eq:fitScaling2} from finite-size scaling are smaller than the marker size.
The exact solution for the $m/g=0$ case, $\omega_1(m/g=0)=1/\sqrt{\pi}\approx0.5641895$ is indicated
with a star.
Displayed are the fits for the whole interval $x\in[20,\,500]$ (dashed line), including up to quadratic terms in $1/\sqrt{x}$, and 
the same fit for $[30,\,300]$ (solid line), but they are practically indistinguishable at the scale of the plots.
The final values for the binding energies, $\frac{M_V}{g}$, and their errors (see Appendix \ref{sec:appendix}) are displayed under the corresponding plots.
}
\label{fig:contVec}
\end{figure}

\begin{figure}[floatfix]
\begin{minipage}[b]{.4\columnwidth}
\subfigure[~Ê$m/g=0$,  $\frac{M_S}{g}=1.1279(12)$]{
\includegraphics[height=.8\columnwidth]{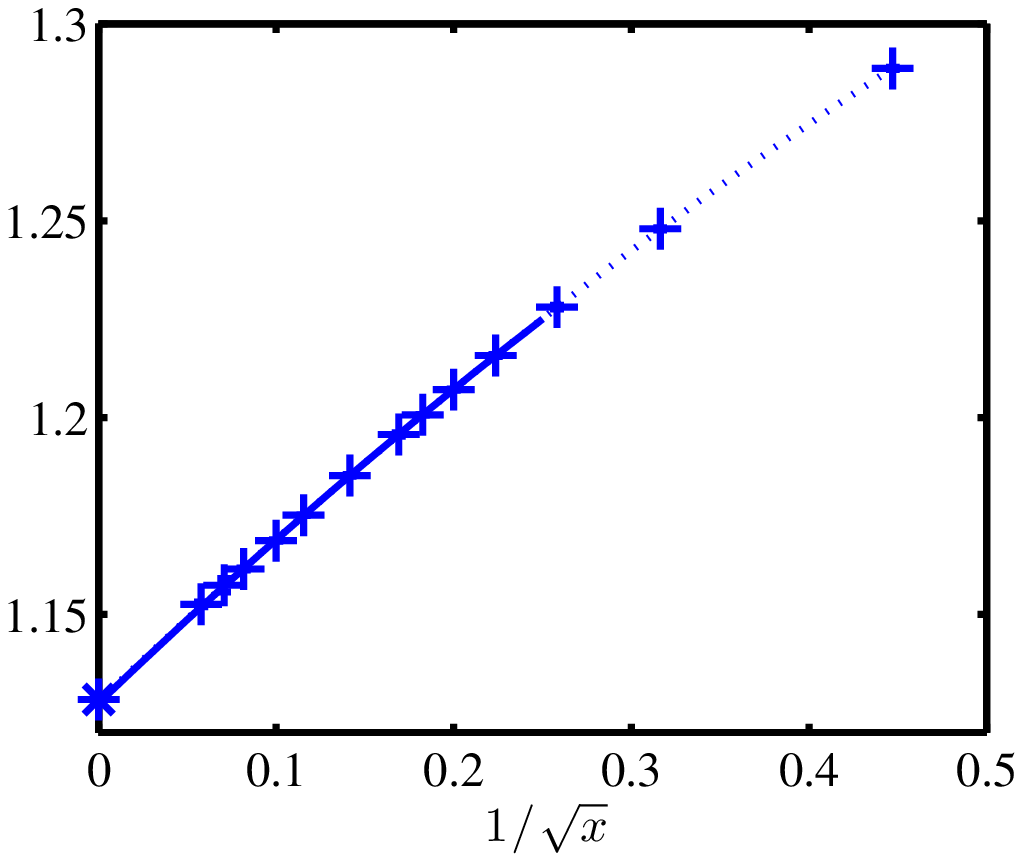}
}
\\
\subfigure[~Ê$m/g=0.125$, $\frac{M_S}{g}=1.2155(28)$]{
\includegraphics[height=.8\columnwidth]{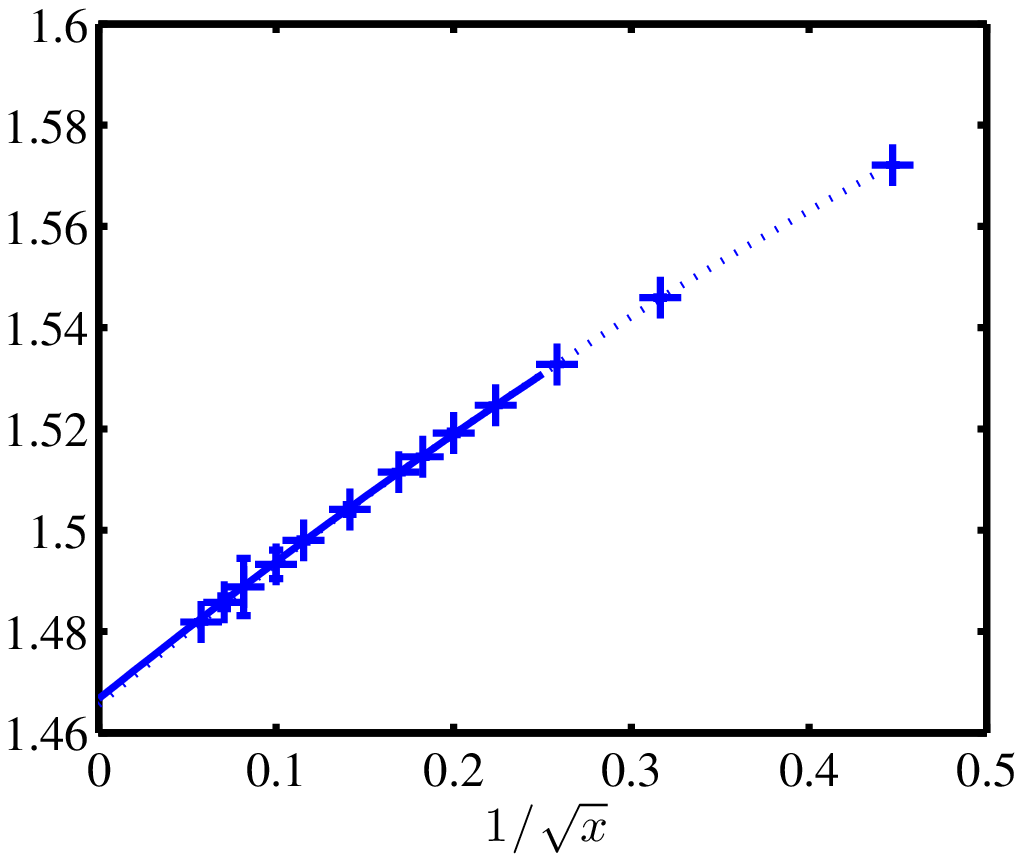}
}
\end{minipage}
\hspace{.1\columnwidth}
\begin{minipage}[b]{.4\columnwidth}
\subfigure[~Ê$m/g=0.25$, $\frac{M_S}{g}=1.2239(22)$]{
\includegraphics[height=.8\columnwidth]{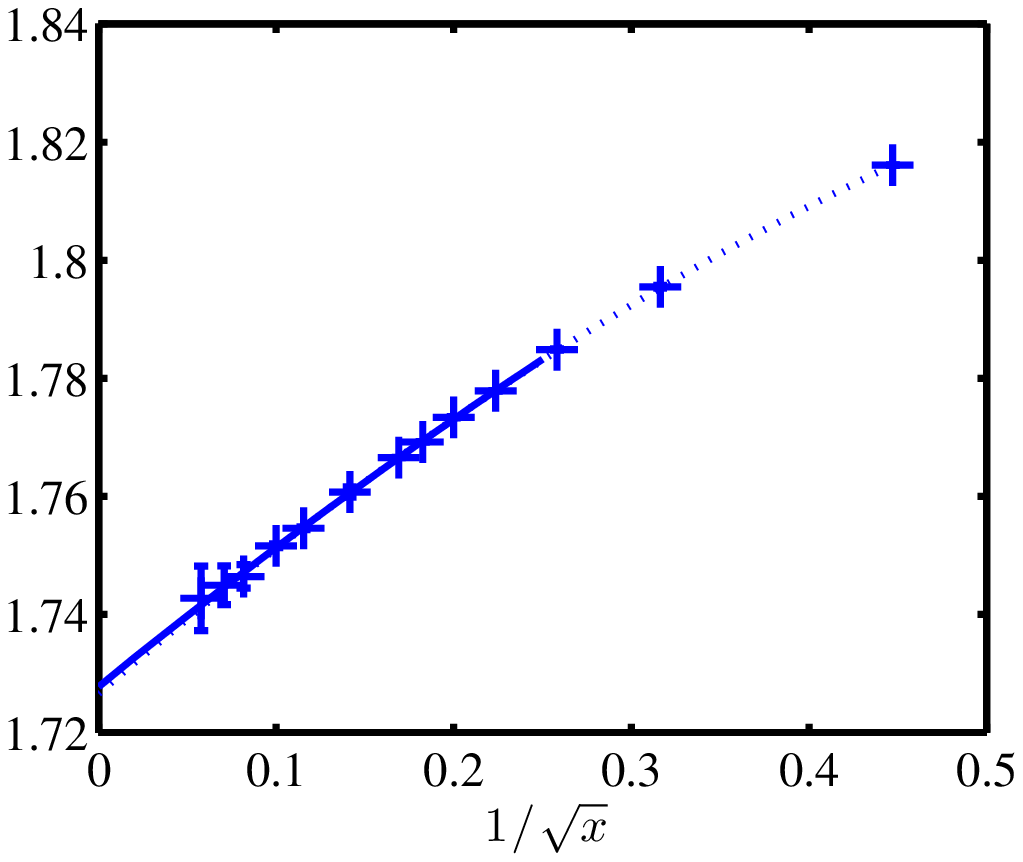}
}
\\
\subfigure[ ~Ê$m/g=0.5$, $\frac{M_S}{g}=1.1998(17)$]{
\includegraphics[height=.8\columnwidth]{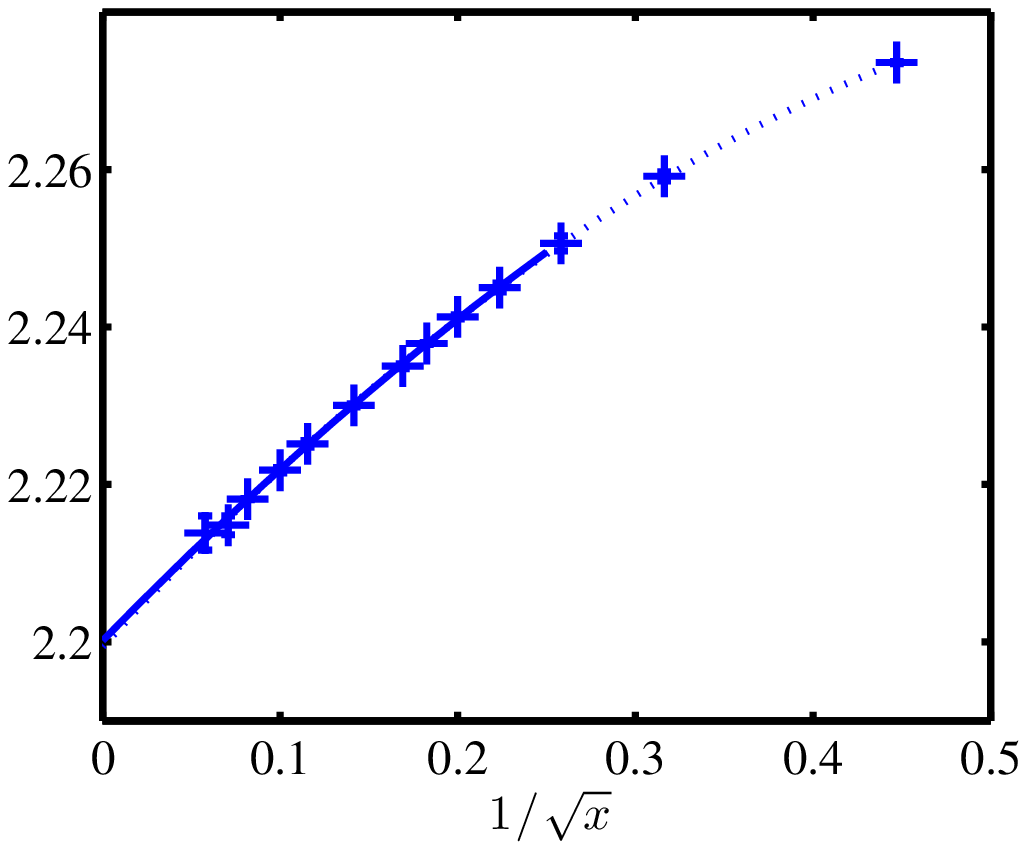}
}
\end{minipage}
\caption{
Energy gap for the scalar state, $\omega_2$,
 as a function of $1/\sqrt{x}$.
The error bars correspond to the uncertainties of the finite-size scaling fits~\eqref{eq:fitScaling2}.
The star on the vertical axes in the $m/g=0$ plot shows the exact solution
$\omega_2(m/g=0)=2/\sqrt{\pi}\approx1.12838$.
Fits for the whole range of $x$ (cubic) and for a smaller interval ($x\in[15,\, 150]$) 
are shown (respectively dashed and solid lines).
The errors for large values of $x$ are in this case much more significant, because they require 
solving larger systems, for some of which the maximum bond dimension reached was not large enough. 
The final values for the binding energies, $\frac{M_S}{g}$, and their errors (computed as described in Appendix \ref{sec:appendix}) are displayed under the corresponding plots.
}
\label{fig:contSca}
\end{figure}

Figures~\ref{fig:contGS} (ground state), ~\ref{fig:contVec} (vector) and ~\ref{fig:contSca} (scalar)
show all the values in the thermodynamic limit as a function of $1/\sqrt{x}$,
for each of the mass values,
and illustrate the fits that produce the continuum limit.

As our final result 
we obtain for the ground state energy density in the massless case the value $\omega_0=-0.318338(22)(24)$, to be
compared to the exact value, $-1/\pi\approx-0.3183099$.
The first error includes the propagated error from infinite $D$ and $N$
extrapolations, as well as the systematic error from choosing the fitting interval, while the second
error is the difference between results from cubic and quartic fits in $1/\sqrt{x}$.
Note that the latter error is important only in the case of the ground state energy -- for the
scalar and vector mass gaps it is negligible with respect to the former error, as the data are very well
described by polynomials only quadratic in $1/\sqrt{x}$.
Our result for the ground state energy in the massive cases is compatible with the massless one,
and independent of $m/g$, as expected \cite{hamer82}.

The results for the vector binding energies, $\frac{M_V}{g}:=\omega_1-2m/g$, are shown in the following table
for each value of $m/g$ studied, 
together with the  
DMRG estimates \cite{byrnes02dmrg} 
 for comparison.
In the massless case, the exact value is also displayed.

\begin{center}
\begin{tabular}{|c|c|c|c|}
\hline
&
\multicolumn{3}{|c|}{Vector binding energy} \\
\hline
$m/g$  & MPS with OBC & DMRG result \cite{byrnes02dmrg} & exact \\
\hline
0 & 0.56421(9)&0.5642(2)& 0.5641895\\
\hline
0.125 & 0.53953(5) & 0.53950(7) & -\\
\hline
0.25 & 0.51922(5) & 0.51918(5) & - \\
\hline
0.5 & 0.48749(3) & 0.48747(2) & - \\
\hline

\end{tabular}
\end{center}

For the scalar, the most precise results found in the literature for the binding energy, $\frac{M_S}{g}:=\omega_2-2m/g$ in the massive case, correspond to the strong coupling
expansion \cite{sriganesh2000}.
 Again, we show them together with our best fit and the exact value for the massless case in the following table.

\begin{center}
\begin{tabular}{|c|c|c|c|}
\hline
&
\multicolumn{3}{|c|}{Scalar binding energy} \\
\hline
$m/g$  & MPS with OBC & SCE result \cite{sriganesh2000} & exact \\
\hline
0 & 1.1279(12) & 1.11(3) &1.12838 \\
\hline
0.125 & 1.2155(28) & 1.22(2) & - \\
\hline
0.25 & 1.2239(22) & 1.24(3) & - \\
\hline
0.5 & 1.1998(17) & 1.20(3) & - \\
\hline

\end{tabular}
\end{center}

\begin{figure}[floatfix]
\begin{minipage}[b]{.4\columnwidth}
\subfigure[~Ê$m/g=0$]{  
\includegraphics[height=.8\columnwidth]{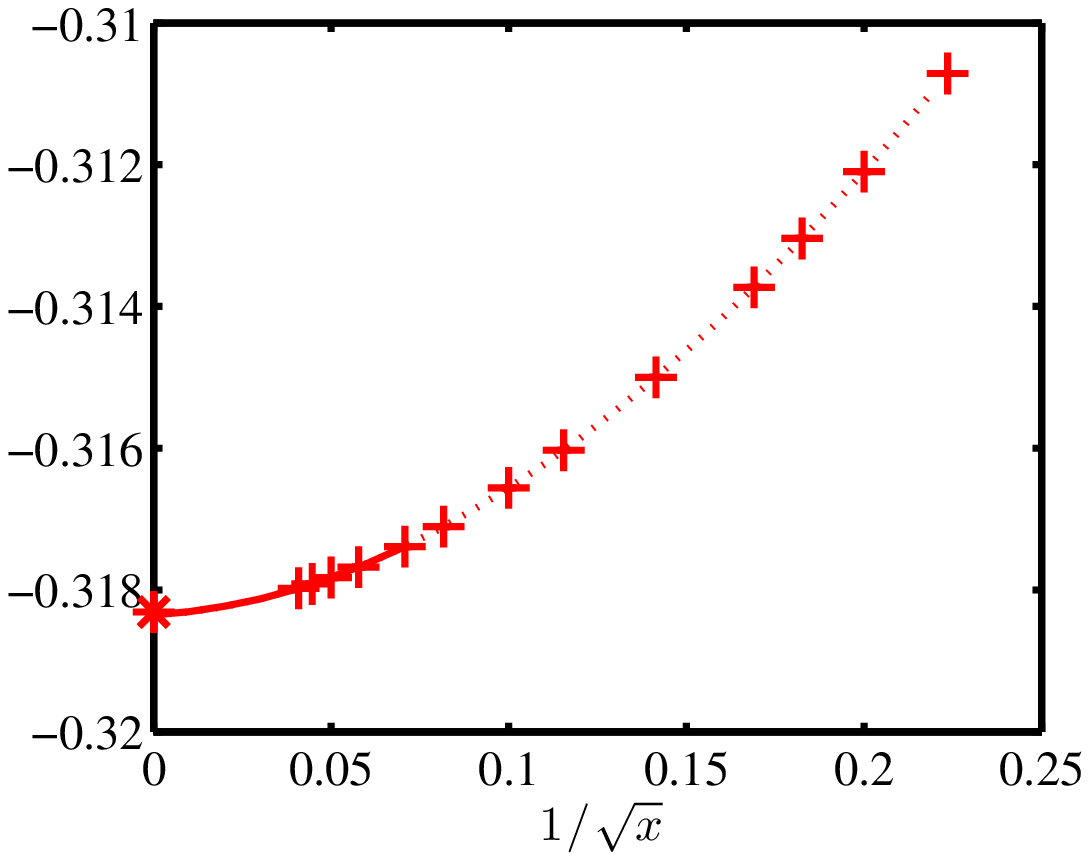}
}
\\
\subfigure[~Ê$m/g=0.125$]{ 
\includegraphics[height=.8\columnwidth]{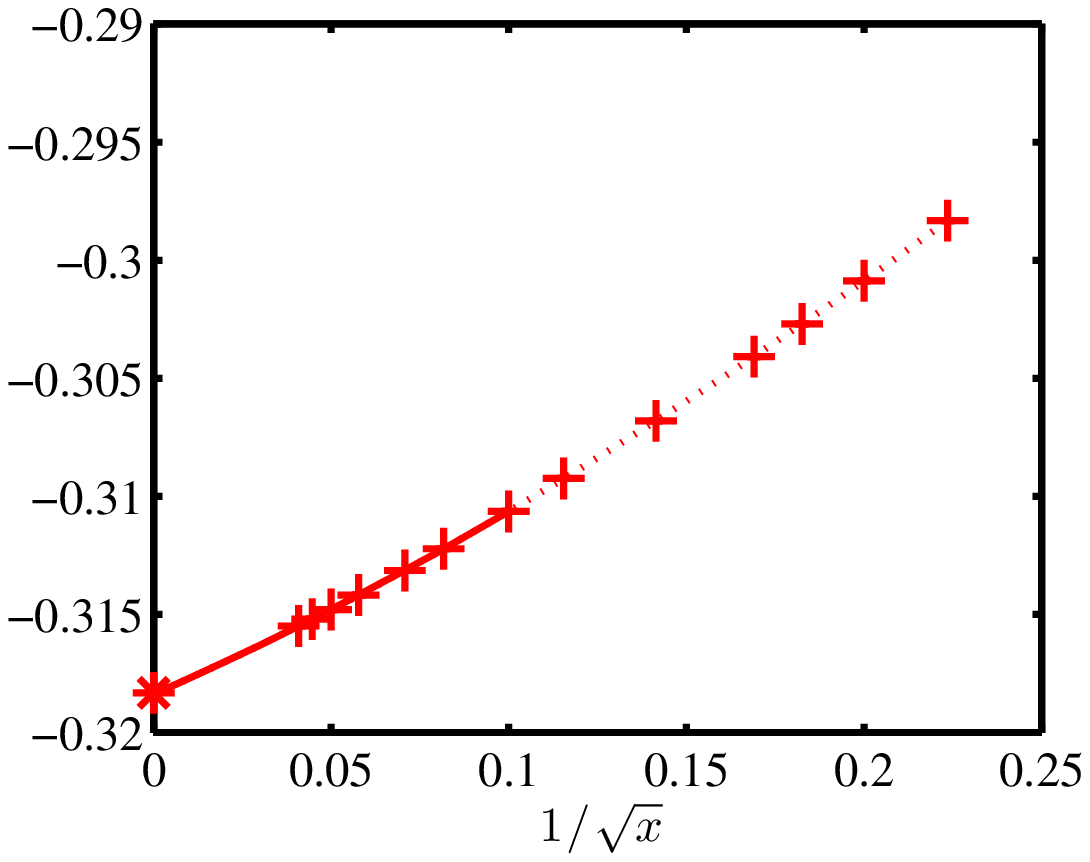}
}
\end{minipage}
\hspace{.1\columnwidth}
\begin{minipage}[b]{.4\columnwidth}
\subfigure[~Ê$m/g=0.25$]{ 
\includegraphics[height=.8\columnwidth]{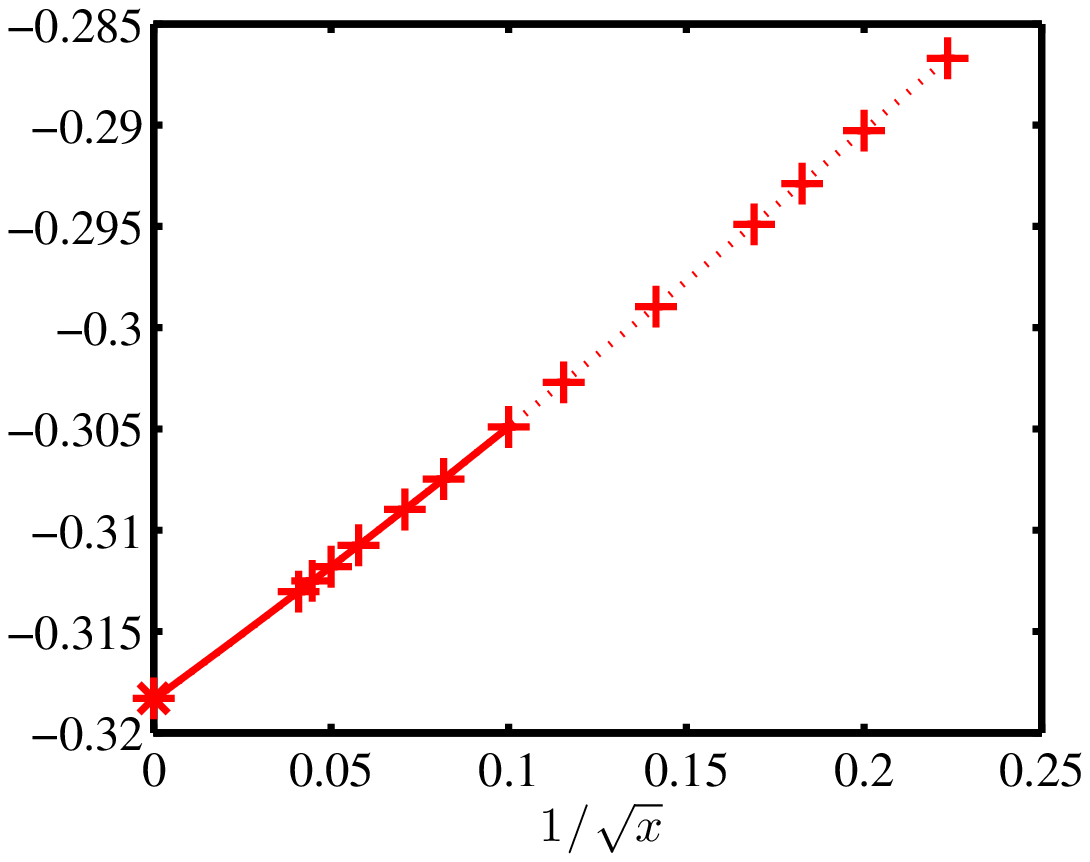}
}
\\
\subfigure[~Ê$m/g=0.5$]{ 
\includegraphics[height=.8\columnwidth]{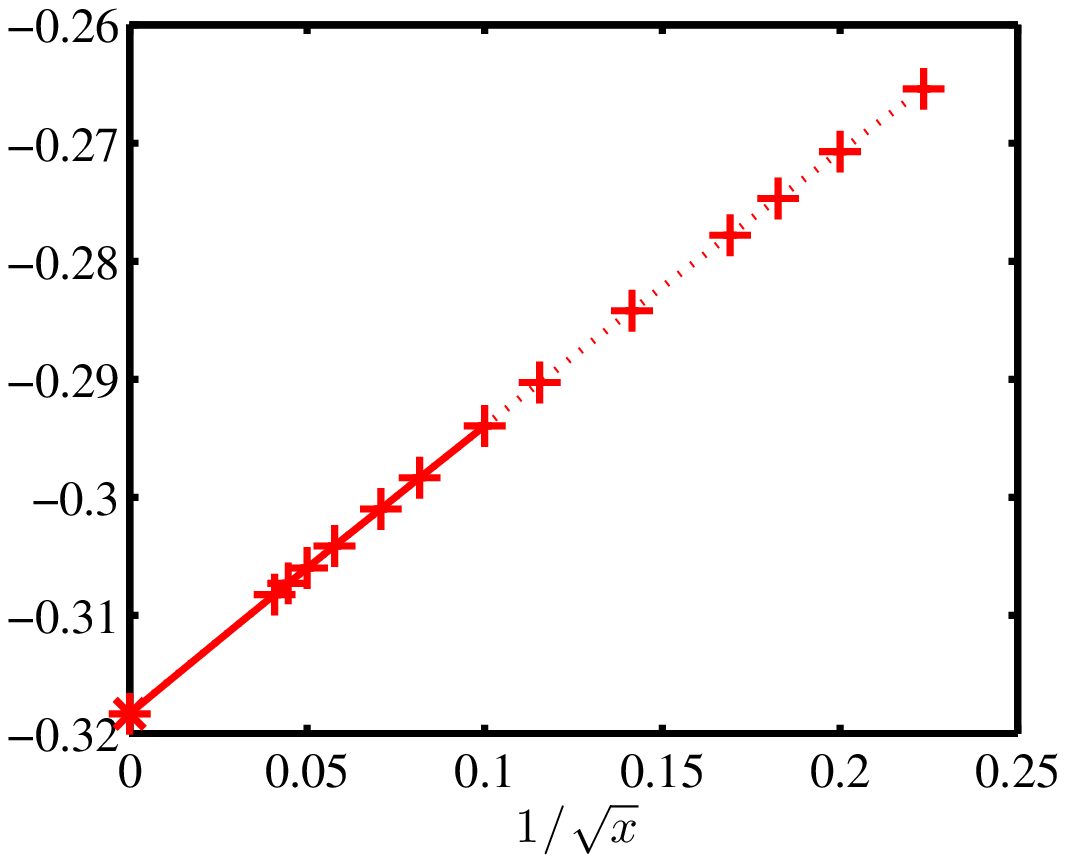}
}
\end{minipage}
\caption{
Ground state energy density as a function of $1/\sqrt{x}$.
The error of each point (smaller than the markers) reflect the uncertainties of the linear fits~\eqref{eq:fitScaling1},
and the star on the vertical axes indicates the (exact) solution for the massless theory,
$\omega_0=-1/\pi\approx-0.31831$.
The dashed line corresponds to the fit of the whole computed range, $x\in[15,\,600]$, to a cubic function in $1/\sqrt{x}$.
Excluding small values $x<50$ produces a better fit, as shown by the solid line, for the interval $x\in[100,\, 600]$.
Our final estimate (see Appendix \ref{sec:appendix}) for the massless case is
$\omega_0=-0.318338(22)(24)$,
and for all other masses it is compatible with this value.
}
\label{fig:contGS}
\end{figure}

One of the factors explaining the lower precision attained in the scalar calculation
 as compared to the vector results 
is the longer time required to reach the scalar state with the MPS algorithm for excited states. 
As discussed in section \ref{sec:method} the computational cost scales like the square
of the required number of levels.
Since in many instances the first scalar candidate appears at level $7-8$ or above,
this represents a cost over $50$ times larger
than in the case of the ground state.
We have thus opted for a trade-off between precision and efficiency, and  
applied for the scalar 
 a  less demanding convergence criterion than 
 for  the vector or the ground state, what also translates in 
 less precision in the final results.
For the same reason,  the largest bond dimensions  are in some cases not available in the scalar 
calculations, which leads to somewhat less precise finite-size scaling.

\section{Discussion}
\label{sec:discussion}

We have successfully employed open boundary MPS to compute the ground state and several
excited levels of
the lattice Schwinger model, using a staggered formulation, and for various values of the fermion mass.
Although in the physical subspace in which we work the model contains long range interactions which do not decay with the distance, we have found that
MPS produce a very good approximation not only to the ground state, but also to higher excited levels, 
and we are able to reach precisions comparable to those available from other techniques,
or 
in some cases
even slightly better.
Additionally, the precision we reach is not extremely sensitive to the value of the mass, what further
points to the usefulness of TN techniques for the non-perturbative parts of the parameter space.

In order to determine the mass gaps, we have also extended the MPS tools to approximate excited states, and we have proposed a 
modified algorithm that is efficient and allows us to approximate more than ten excited states in chains of hundreds of sites.
These techniques provide an ansatz for the targeted states, so that not only the energies, but also other observables can be precisely determined~\cite{banuls13cond}.

Our results validate the applicability of tensor network techniques, in particular MPS, for lattice gauge theory problems.
We have obtained very precise results,
even though the open boundary MPS cannot represent 
states with well-defined momentum, a problem that a 
TI ansatz might likely overcome. 

The long term goal would be to generalize these studies to higher dimensional systems,
getting in this way closer to the target of lattice QCD studies.
The MPS ansatz employed in this work is one-dimensional, and its applicability to higher dimensional 
problems is limited, a fact well understood in terms of its entanglement content.
Nevertheless, the entanglement structure in the MPS   
has a natural generalization to two and higher dimensions
in the PEPS ansatz~\cite{verstraete04peps,verstraete08algo}.
Algorithms exist for the ground state search and for simulating the
evolution also in the case of PEPS,
and the ansatz can directly be applied to fermionic systems~\cite{kraus10fpeps,corboz10fpeps,pineda10fpeps}.
Although the higher computational costs limit the nowadays feasible 
system sizes and bond dimensions,
PEPS are good candidates to study higher dimensional lattice gauge problems.
Our MPS study is a first step to assess the feasibility of TNS to describe the relevant physics in this
kind of problems.

One of the main advantages of the MPS and other TNS methods is that they easily allow us to attack other problems which are more complicated 
for standard lattice techniques. In particular, it is straightforward to include a chemical potential term in the Hamiltonian, which is interesting in many-flavour
Schwinger models. 
Additionally, we could perform a similar study in a more complicated lattice theory, in particular
with a  non-Abelian symmetry.
Thermal equilibrium properties
can be easily studied with TNS, and it
 is also possible to simulate real time evolution, which opens the door to out-of-equilibrium questions.
Thus even if tensor networks are presently not competitive with standard Monte Carlo
techniques for 4D quantum field theories, they might allow us to address problems that are not amenable for customary lattice field
theory techniques. 


\acknowledgments

\sloppy
This work was partially funded by the EU through SIQS grant (FP7 600645).
K.C. was supported by Foundation for Polish Science fellowship ``Kolumb''.
This work was supported in part by the DFG Sonderforschungsbereich/Transregio SFB/TR9. 
K.J. was supported in part by the Cyprus Research Promotion
Foundation under contract $\Pi$PO$\Sigma$E$\Lambda$KY$\Sigma$H/EM$\Pi$EIPO$\Sigma$/0311/16.

\appendix
\section{Continuum limit and error analysis}
\label{sec:appendix}
As we have described in the main body of the text, we have a set of data corresponding to a given bond
dimension $D$, lattice size $N$ and coupling $x$. Having performed extrapolations to infinite $D$ and
$N$, we are left with the dependence of our observables on $x$ and we want to take the continuum limit
$x\rightarrow\infty$. 
The leading order cut-off effects are $\mathcal{O}(1/\sqrt{x})$, but we are sensitive also to
higher-order corrections ($\mathcal{O}(1/x)$, $\mathcal{O}(1/x^{3/2})$, $\ldots$).
Therefore, in general, our continuum extrapolated result depends on the chosen fitting interval in $x$.
To take this systematic uncertainty into account, we perform the following error analysis procedure,
similar to 
those employed in data analysis for Lattice QCD (see
e.g. Refs.~\cite{Baron:2009wt,Durr:2010aw}).

Let us assume that,
 for a fixed value of $m/g$,
 we have a set of data for the chosen observable, $O$: 
 $\{(x_i,\,O_i,\,\Delta O_i)\}_{i=1}^n$,
  for $n$ values of $x$, 
 such that $O_i$ is the observable evaluated at a finite lattice spacing
corresponding to coupling $x_i$, and $\Delta O_i$ is its respective error (originating both from
extrapolations to infinite $D$ and $N$).
We choose some minimal number of points ($n_{\mathrm{min}}$) and we 
fit every possible subset of  
 $p$ consecutive data points,
with $p\geq n_{\mathrm{min}}$,
to a given polynomial function $g(x)$. 
Each such fit gives us an estimate of the continuum value of the chosen observable, $O^{\mathrm{c}}$ -- we denote
these continuum values by $O_{\alpha}^{\mathrm{c}}$, where $\alpha=1,\,\ldots,\,M_{\mathrm{fits}}$.
Each fit has its respective value of $\chi^2$, denoted by $\chi^2_{\alpha}$:
\begin{equation}
 \chi^2_{\alpha}=\sum_{i:\ x_i\in{\rm fit\,}{\alpha}} \frac{\left(g(x_i)-O_i\right)^2}{(\Delta O_i)^2}
\end{equation}
where 
 the sum
runs over all data points entering the 
fit labelled by $\alpha$. 

Our preferred value for the observable in the continuum limit is then defined
as the \emph{median} of the distribution of estimators $O_{\alpha}^{\mathrm{c}}$ weighted by 
 $\exp(-\chi^2_{\alpha}/N_{\rm d.o.f.})$, where $N_{\rm d.o.f.}$ is the number of degrees
of freedom.

To estimate this quantity, the fits are reordered such that for all ${\alpha}$, $O_{\alpha}<O_{{\alpha}+1}$.
We then define the cumulative distribution function $f_{\alpha}$ (${\alpha}=1,\,\ldots,\,M_{\mathrm{fits}}$):
\begin{equation}
 f_{\alpha}=\frac{\sum_{\kappa=1}^{\alpha} \exp(-\chi^2_{\kappa}/N_{\rm d.o.f.})}{\sum_{\kappa=1}^{M_{\mathrm{fits}}} \exp(-\chi^2_{\kappa}/N_{\rm
d.o.f.})}.
\end{equation}  
Our preferred value for the considered observable in the continuum limit is then defined as the value of
$O_{\alpha}$ for which $\alpha$ is the smallest number that satisfies $f_{\alpha}\geq0.5$.
 The corresponding error (the systematic error from the choice of the fitting range) is computed as
$0.5(O^+ - O^-)$, where $O^+$ is the value of $O_{\alpha}$ corresponding to the smallest ${\alpha}$ such that
$f_{\alpha}\geq0.8415$ and $O^-$ is the value of $O_{\alpha}$ corresponding to the smallest ${\alpha}$ such that
$f_{\alpha}\geq0.1585$.
This definition of the error is motivated by the fact that in the limit of an infinite number of fits,
the weighted distribution will be Gaussian and  
this definition of the error 
will correspond to the width of such
Gaussian distribution.

\bibliography{MPSSchwinger}

\end{document}